\documentclass[conference]{IEEEtran}

\ifCLASSINFOpdf
\else
\fi

\usepackage[utf8]{inputenc}
\usepackage{graphicx} %
\usepackage{algorithm,algorithmicx,verbatim,color}
\usepackage[noend]{algpseudocode}
\usepackage{amsmath, amssymb}
\usepackage{cleveref}          %
\usepackage{booktabs}
\usepackage{url}

\usepackage{multirow}
\usepackage{subcaption}
\usepackage{bm}
\usepackage{xcolor} %

\newcommand{\commentblue}[1]{\texttt{\textcolor{gray}{// #1}}}

\newcommand{\g}{\boldsymbol{g}}  %
\newcommand{\z}{\boldsymbol{z}}  %
\newcommand{\w}{\mathbf{w}}      %
\newcommand{\G}{\mathbf{G}}      %

\newtheorem{property}{Property}

\usepackage{pifont}

\newcommand{\yes}{\ding{108}}       %
\newcommand{\no}{\ding{109}}            %

\newcommand{\ex}[2]{{\ifx&#1& \mathbb{E} \else \underset{#1}{\mathbb{E}} \fi \left[#2\right]}}
\newcommand{\pr}[2]{{\ifx&#1& \mathbb{P} \else \underset{#1}{\mathbb{P}} \fi \left[#2\right]}}
\newcommand{\exc}[3]{{\ifx&#1& \mathbb{E} \else \underset{#1}{\mathbb{E}} \fi \left[ #2 \middle| #3 \right]}}
\newcommand{\prc}[3]{{\ifx&#1& \mathbb{P} \else \underset{#1}{\mathbb{P}} \fi \left[ #2 \middle| #3 \right]}}
\newcommand{\var}[2]{{\ifx&#1& \mathbb{V} \else \underset{#1}{\mathbb{V}} \fi \left[#2\right]}}

\newcommand{\nope}[1]{}

\newtheorem{theorem}{Theorem}

\newtheorem{definition}[theorem]{Definition}

\newcommand{\todo}[1]{\textcolor{red}{TODO}}
\newcommand{\ours}[1]{\texttt{UniAud}}
\newcommand{\ourss}[1]{\texttt{UniAud++}}

\begin{document}
\title{UniAud: A Unified Auditing Framework for High Auditing Power and Utility with One Training Run}

\author{
\IEEEauthorblockN{Ruixuan Liu}
\IEEEauthorblockA{Emory University\\
ruixuan.liu2@emory.edu}
\and
\IEEEauthorblockN{Li Xiong}
\IEEEauthorblockA{Emory University\\
lxiong@emory.edu}
}

\maketitle

\begin{abstract}
Differentially private (DP) optimization has been widely adopted as a standard approach to provide rigorous privacy guarantees for training datasets. DP auditing verifies whether a model trained with DP optimization satisfies its claimed privacy level by estimating empirical privacy lower bounds through hypothesis testing.
Recent O(1) frameworks improve auditing efficiency by checking the membership status of multiple audit samples in a single run, rather than checking individual samples across multiple runs.
However, we reveal that there is no free lunch for this improved efficiency: data dependency and an implicit conflict between auditing and utility impair the tightness of the auditing results.
Addressing these challenges, our key insights include reducing data dependency through uncorrelated data and resolving the auditing-utility conflict by decoupling the criteria for effective auditing and separating objectives for utility and auditing.
We first propose a unified framework, \ours~, for data-independent auditing that maximizes auditing power through a novel uncorrelated canary construction and a self-comparison framework. We then extend this framework as \ourss~ for data-dependent auditing, optimizing the auditing and utility trade-off through multi-task learning with separate objectives for auditing and training.
Experimental results validate that our black-box O(1) framework matches the state-of-the-art auditing results of O(T) auditing with thousands of runs, demonstrating the best efficiency-auditing trade-off across vision and language tasks. 
Additionally, our framework provides meaningful auditing with only slight utility degradation compared to standard DP training, showing the optimal utility-auditing trade-off and the benefit of requiring no extra training for auditing.

\end{abstract}

\IEEEpeerreviewmaketitle

\section{Introduction}
As machine learning models gain widespread adoption in sensitive domains, data privacy has emerged as a critical concern. A compelling body of evidence demonstrates that trained models can inadvertently leak private information about their training data—through model gradients~\cite{zhu2019deep}, loss values~\cite{carlini2022membership}, or output predictions~\cite{carlini2021extracting}. 
These findings underscore the urgent need for rigorous privacy guarantees throughout the learning process. \looseness=-1

Differentially private (DP) training, such as DP-SGD~\cite{abadi2016deep}, has become a standard approach for privacy-preserving machine learning.  
According to the DP definition~\cite{dwork2016calibrating}, the probability that an adversary can distinguish between two models trained on datasets differing by a single data point is bounded by the privacy parameters $(\epsilon, \delta)$.  
This inherent guarantee makes DP a natural defense against membership inference attacks (MIAs)~\cite{carlini2022membership, shokri2017membership}, which aim to determine whether a particular individual's data was included in the training set.  
Moreover, DP has been adopted as a foundational component in mitigating other privacy attacks~\cite{carlini2021extracting}.

While DP training has been widely adopted, the strength of its guarantees critically depends on correct implementation.  
To achieve the claimed privacy level, DP training requires random batch sampling, per-sample gradient clipping, and gradient perturbation, where any flawed implementation can result in higher empirical privacy risk~\cite{tramer2022debugging}.  
In contrast to the upper bound provided by DP, DP auditing estimates a lower bound on the actual privacy risk. 
A tight audit implies that the empirical lower bound closely approaches the theoretical upper bound.  
From the analyzer's perspective, DP auditing serves to assess the tightness of the analytical bounds.  
From the practitioner's perspective, it helps uncover implementation flaws, such as incorrect noise injection or faulty gradient clipping, in DP training systems.

Essentially, DP auditing works by formulating the process as a hypothesis test that determines whether to reject the null hypothesis: that the target model satisfies its claimed privacy level at a given confidence level.  
It relies on empirical observations by guessing the membership of an audit sample, in alignment with DP guarantees.  
The original auditing algorithms~\cite{jagielski2020auditing, nasr2021adversary, nasr2023tight} have $O(T)$ complexity, as they require repeatedly training models on neighboring datasets that differ by one sample, for $T$ observations.  
A recent $O(1)$ method~\cite{steinke2023privacy} improves efficiency by randomly including or excluding multiple audit samples for obtaining multiple observations with a single training run.

\noindent\textbf{Challenges in Black-Box O(1) Auditing}. 
Despite recent progress, a critical trade-off persists: $O(T)$-methods~\cite{jagielski2020auditing,nasr2021adversary} achieve tighter empirical lower bounds (i.e., smaller gaps to the theoretical $\epsilon$) by leveraging a greater number of observations, corresponding to $T$ training runs.  
In contrast, $O(1)$-methods~\cite{steinke2023privacy, xiang2025privacy, mahloujifar2024auditing} sacrifice auditing power for scalability, especially when the number of audit samples $m$ in a single training run increases.  
This gap is further exacerbated in black-box settings, where auditors have access only to model outputs (e.g., APIs or prediction interfaces)—a realistic scenario in third-party audits or regulatory compliance.  
For example, black-box $O(1)$ auditing is significantly looser than its white-box counterpart~\cite{steinke2023privacy, xiang2025privacy}, potentially understating actual privacy risks.  
Nevertheless, black-box access is often necessary~\cite{nasr2021adversary, muthu2024nearly} due to practical constraints in commercial deployments or limited system visibility available to auditors.

It remains an open problem to bridge the gap between efficiency and auditing power in black-box $O(1)$ auditing.  
This limitation undermines trust in DP deployments where transparency is limited—precisely the scenarios in which reliable auditing for the training algorithm is most critical.

We note that the challenge of the auditing-efficiency trade-off in $O(1)$ frameworks stems from two key factors.  
On the one hand, auditing power is weakened by data dependencies within real-world audit datasets.  
Correlated audit samples influence each other’s membership inference decisions—not only through interactions during training, but also because existing $O(1)$ auditing algorithms rely on relative ranking to predict membership.  
On the other hand, current $O(1)$ frameworks implicitly introduce an auditing-utility trade-off, complicating the auditing-efficiency dilemma into a trilemma.  
Existing methods~\cite{panda2023differentially} attempt to achieve both meaningful auditing results and strong model utility, but reconciling these two objectives is inherently difficult.

\noindent\textbf{Key Insights.}
Our goal is to improve the auditing power of the efficient $\mathcal{O}(1)$ framework by addressing the aforementioned challenges.  
First, we observe that data dependency manifests not only during training but also in the membership inference phase of existing $\mathcal{O}(1)$ auditing algorithms.  
Thus, our key intuition is to minimize data dependency across both the training and inference stages.  
Second, we find that current DP auditing goals lack clear definitions when simultaneously considering the three aspects of auditing power, efficiency, and utility.  
Our key insight is to disentangle these factors—particularly by decoupling utility and auditing—while preserving the computational efficiency of $\mathcal{O}(1)$ methods.

\noindent\textbf{Our Solutions.}
To minimize data dependency during training, we propose a unified random pair-matching task that encodes the membership of each sample independently. 
This design promotes greater independence in the influence of each audit sample on the model.  
To reduce data dependency during the membership inference phase, we introduce a novel self-comparison-based inference method that removes reliance on relative ranking across samples, thereby making the MIA decision for each audit sample independent.  
To disentangle auditing and utility goals under efficient $O(1)$ computation, we explicitly distinguish between auditing objectives in data-independent and data-dependent settings—a distinction that has not been addressed in prior work.
For data-independent auditing, which aims to evaluate the DP implementation independent of the training data or model, we design a separate auditing dataset and maximize auditing capability under $\mathcal{O}(1)$ efficiency.  
For data-dependent auditing, which targets estimating the empirical privacy risk lower bound for a specific training dataset and model, we consider the triplet trade-offs and resolve the auditing-utility conflict through a decoupled training objective.

We summarize our contribution as follows:
\begin{itemize}
    \item We deeply analyze the challenge in existing O(1) framework, and reveal two fundamental insights for a unified black-box O(1) auditing solution: reducing data dependency and decoupling utility from auditing.
    Thus, we clarify data-independent and data dependent auditing goals and set up for the unified auditing framework, which is also general for different tasks and input formats (e.g., images and text sequences).
    \item To improve the auditing-efficiency trade-off in the data-independent setting, we propose the basic framework \ours~, which addresses the key challenge of data dependency through a novel pair-matching auditing task with synthetic canaries and self-comparison inference.
    \item To optimize the trade-off among auditing, efficiency, and utility in data-dependent auditing, we extend \ours~ to \ourss~ through a novel design that decouples the training objectives of membership encoding (for auditing) and the main task (for utility).
	\item Through extensive evaluation on both image classification and language modeling tasks, we show that \ours~ significantly improves the auditing performance of O(1) methods in the data-independent setting, matching the state-of-the-art O(T) auditing performance, while maintaining O(1) efficiency and saving $\times 10^3$ of training runs compared to O(T) methods; \ourss~ achieves the best overall trade-off compared to existing methods in the data-dependent setting, demonstrating the potential of integrating auditing along with normal training without extra runs.
\end{itemize}

\section{Related Works}
\label{sec:related-work}

\subsection{Theoretical Privacy Auditing}
Privacy auditing~\cite{steinke2023privacy,jagielski2020auditing, nasr2021adversary, nasr2023tight} has emerged as a crucial tool for validating the empirical privacy guarantees of differentially private algorithms.
For example, it helps to identify implementation flaws in DP systems, such as incorrect noise calibration or gradient clipping mechanisms~\cite{tramer2022debugging}. 
The original work~\cite{jayaraman2019evaluating} demonstrated how standard membership inference attacks~\cite{shokri2017membership} could evaluate privacy analysis algorithms, while ~\cite{jagielski2020auditing} introduced worst-case poisoning examples to stress-test privacy bounds.
For example, the empirical risk can be lower bounded through the membership inference performance $\epsilon \geq \ln(\text{TPR}/\text{FPR})$~\cite{steinke2023privacy} given ensured statistical validity.

Recent advances in DP auditing have progressed along several key dimensions:
1) \textbf{Tighter Privacy Analysis:}
The work of~\cite{nasr2021adversary} established that DP-SGD's analysis is tight within its assumed worst-case threat model.
~\cite{nasr2023tight} and~\cite{maddock2022canife} developed tighter auditing by exploiting the iterative nature of DP-SGD via auditing individual steps.
Other approaches improved statistical estimation through Log-Katz confidence intervals~\cite{lu2022general} or Bayesian methods~\cite{zanella2023bayesian}. 
2) \textbf{Efficiency Improvements:} 
~\cite{andrew2023one} proposed a method to audit excluded data points without algorithm re-execution, while~\cite{pillutla2023unleashing} re-uses training runs over overlapping dataset pairs to improve efficiency, though still requiring multiple runs.
The recent work~\cite{steinke2023privacy} significantly improves the auditing efficiency via the heuristic of including/excluding multiple samples in one training run and providees a theoretical bound.
Recent works~\cite{mahloujifar2024auditing, xiang2025privacy} follow the similar framework further improves the tightness by leveraging f-DP definitions.

Compared with these works, our work is based on the most general O(1) framework~\cite{steinke2023privacy}, and improvements via tighter statistical or other DP definitions are orthogonal to ours.

\subsection{Empirical Privacy Auditing}
Not limited to auditing a theoretical lower bound for DP trained model, the general term of privacy auditing includes membership inference attack, quantifiable data proving and membership encoding.
The success of O(1)~\cite{steinke2023privacy} formally proves that multi-example membership inference can be used for auditing purposes, while it also claims that it is non-trivial to leverage advanced MIA such as considering the sample hardness~\cite{carlini2022membership}.
Similarly, the recent advanced MIA with privacy backdoor~\cite{liu2024precurious, wen2024privacy} has been leveraged to improve O($T$).
Quantifiable data proving~\cite{huang2024general, tongmuch} is an instance of membership inference attack that only infers the existence of data from one owner with a certain confidence.
For example, \cite{huang2024general} creates two versions of the protected dataset with small perturbation and formulate the null hypothesis as whether the relased version has been used to train.
Membership encoding~\cite{song2020membership,chen2024method} is another proactive form of empirical auditing by manipulating the training process to force the model to encode membership status of its training dataset.
However, these works do not directly assist DP auditing.

Compared with above empirical privacy auditing works, our work solves the problem of how to leverage sample hardness for DP  auditing, and leverages insights from data proving and membership encoding for tighter theoretical DP auditing.

\section{Preliminaries}
We will introduce background of DP training, DP auditing and our threat model for black-box O(1) auditing.
We summarize necessary notations in \Cref{tab:notation}.
\begin{table}[t]
\caption{Notation Summary}
\label{tab:notation}
\centering
\resizebox{0.98\linewidth}{!}{
\begin{tabular}{@{}lll@{}}
\toprule
\textbf{Symbol} & \textbf{Description} & \textbf{Example/Definition} \\
\midrule
$\epsilon,\delta$ & Privacy parameters & $(\epsilon,\delta)$-DP guarantee \\
$R_g$ & Gradient clipping norm & $\min\{\frac{R_g}{\|\g_i\|},1\}\g_i$ \\
$\z$ & Noise vector & $\z \sim \mathcal{N}(0,I)$ \\
$\sigma$ & Noise multiplier & Scales Gaussian noise in DP-SGD  \\
$\g_i$ & Per-example gradient & $\nabla_\theta l(\hat{p}(y|x_i,\theta), y_i)$ \\
\midrule
$D$ & Dataset & $\{(x_i,y_i)\}_{i=1}^n$ \\
$\mathcal{X}$ & Input space & Feature domain (e.g., images, text) \\
$\mathcal{Y}$ & Output space & Label domain (e.g., classes, tokens) \\
$C$ & Output space size & $y\in\mathcal{Y}, C:=|\mathcal{Y}|$ \\
$B$ & Mini-batch & Number of samples in one batch $\mathcal{B}$ \\
$\theta$ & Model parameters & $\theta \in \mathbb{R}^d$ \\
$f_\theta$ & Model function & Mapping $x \mapsto f_\theta(x)$ \\
$\hat{p}(y|x,\theta)$ & Prediction distribution & $\text{softmax}(f_\theta(x))$ \\
\hline
$T$ & Audit trials & Number of hypothesis tests \\
$m$ & Audit data size & $m > 0$ \\
$n$ & Training data size & $n \geq m$ \\
$\alpha$ & Significance level & Type I error probability (e.g., 0.95) \\
$\epsilon_O$ & Optimal estimation & Assuming perfect MIA given $m$ \\
\bottomrule
\end{tabular}
}
\end{table}

\subsection{Differentially-Private Training}\label{sec:dpsgd}
Differential privacy (DP) provides a rigorous mathematical framework for limiting the influence of any single training example on the output of an algorithm. Formally, a randomized algorithm is said to satisfy DP if the inclusion or exclusion of any single data point does not significantly change the distribution of its outputs. This is captured by the following definition.

\begin{definition}[Differential Privacy]
A randomized algorithm $\mathcal{A}$ satisfies $(\epsilon, \delta)$-differential privacy if for any two datasets $D$ and $D'$ differing in at most one entry, and for all measurable subsets $S$ of the output space of $\mathcal{A}$, we have
\[
\Pr[\mathcal{A}(D) \in S] \leq e^\epsilon \Pr[\mathcal{A}(D') \in S] + \delta.
\]
\end{definition}

DP training algorithms, such as Differentially Private Stochastic Gradient Descent (DP-SGD), ensure the above definition by carefully injecting noise during the training process. 
Specifically, DP-SGD modifies standard SGD by clipping per-sample gradients to limit individual influence and then adding calibrated Gaussian noise to the aggregated gradients before updating the model parameters. 
This process guarantees that the final model's behavior does not rely heavily on any single training example, thereby satisfying $(\epsilon, \delta)$-differential privacy, which we also call it analytical privacy budget or the worst-case privacy upper bound.

For example at the iteration $t$, SGD optimization updates model as $\w_{t+1} = \w_t - \eta \cdot \G_t$, and DP-SGD privatizes the updates: 
\begin{align}
\text{Non-DP:}\quad \G_t &= \frac{1}{|\mathcal{B}|}\sum_{i \in \mathcal{B}} \g_i \\
\text{DP-SGD:}\quad \G_t &= \underbrace{\frac{1}{|\mathcal{B}|}\sum_{i \in \mathcal{B}} \min\left\{\frac{R_g}{\|\g_i\|},1\right\}\g_i}_{\text{clipped gradients}} + \underbrace{\sigma R_g \z}_{\substack{\text{noise scaled to} \label{eq:dpsgd} \\ \text{clipping norm }R_g}}
\end{align}
$\g_i$ is the gradient of the loss with respect to the $i$-th sample in the mini-batch $\mathcal{B}$, $R_g$ is the clipping norm, $\sigma$ is the noise multiplier, and $\z \sim \mathcal{N}(0, I)$ is standard Gaussian noise.

In general, the DP training algorithm $\mathcal{T}(D, \theta_0|\epsilon, \delta)$ takes the private dataset $D$ and the initialized model $\theta_0$ as input and outputs a trained model $\theta$, which is claimed to satisfy $(\epsilon, \delta)$-DP guarantee that is independent on either $D$ or $\theta_0$.

\subsection{Privacy Auditing for DP Training}
Differential privacy by its definition bounds the strongest adversary's capability to infer if $\mathcal{T}$ is trained on $D$ or $D^\prime$ differing by one sample.
Without loss of generality, suppose guessing $D^\prime$ when the model is trained on $D$ as false positive, the false positive $\bm{\alpha}$ and false negative $\bm{\beta}$ are bounded by the privacy region~\cite{kairouz2015composition}:
\begin{align}
\mathcal{R}(\epsilon, \delta) 
&= \{ (\bm{\alpha}, \bm{\beta}) \mid 
\quad \bm{\alpha} + e^{\epsilon} \bm{\beta} \geq 1 - \delta \land e^{\epsilon} \bm{\alpha} + \bm{\beta} \geq \nonumber 1 - \delta \\
& \quad \land\ \bm{\alpha} + e^{\epsilon} \bm{\beta} \leq e^{\epsilon} + \delta \land e^{\epsilon} \bm{\alpha} + \bm{\beta} \leq e^{\epsilon} + \delta \} \label{eq:region}
\end{align}
Ideally, $\epsilon$ can be computed by fixing $\delta$ given $\bm{\alpha}$ and $\bm{\beta}$ via \Cref{eq:region}.
However, since the minimum possible values of $\bm{\alpha}$ and $\bm{\beta}$ are hard to be computed in closed form, empirical estimates are necessary.

The essence of auditing algorithm is a hypothesis test to distinguish the model is trained on $D$ or $D^\prime$ given multiple empirical observations.
In the $i^\text{th}$ observation, the auditor makes a guess on the membership status with respect to a sample $z_i$ as $\hat{S}_i$ given the model $\theta$ trained by $\mathcal{T}$:
\begin{align}
    \hat{S}_i = f_\text{MIA}(I_i), \text{where } I_i=f_\text{SCORE}(z_i, \theta)
\end{align}
A widely used scoring function is the negative loss $f_\text{SCORE}(z_i) = -\mathcal{L}(z_i, \theta)$, which higher score indicates lower loss and higher chance to be included in training dataset.
Thus, the membership inference function $f_\text{MIA}(I_i)$ output binary prediction given a threshold on scores.

Given guesses and the ground truth membership $S$ for multiple observations, it derives the confidence intervals for $\bm{\alpha}$ and $\bm{\beta}$ and then convert to transfer the lower bound of $\epsilon$ given $\delta$ at a certain confidence level: 
\begin{align}
    \epsilon_T = f_\text{EST}(S, \hat{S} |\delta, 1-\beta)
\end{align}

For example, after obtaining the empirical lower and upper bounds using binomial proportion confidence interval for $\bm{\alpha}$ and $\bm{\beta}$.
\cite{nasr2021adversary} uses the Clopper-Pearson method to find the upper bound of errors and derives the empirical lower bound of $\epsilon$ as:
\begin{align}
    \epsilon_L = \max \left\{ 
    \ln \left( \frac{1 - \bar{\bm{\alpha}} - \delta}{\bm{\beta}} \right),\,
    \ln \left( \frac{1 - \bar{\bm{\beta}} - \delta}{\bm{\alpha}} \right),\,
    0
    \right\} \label{eq:nasr_lb}
\end{align}
which is shown to be tight in O($T$) auditing for malicious dataset attack with $D_{\backslash z_i}=\varnothing$ which matches the worst-case in DP definition.
Recent works improve $f_\text{EST}$ tighter lower bound by using different confidence intervals~\cite{lu2022general}, Bayesian techniques~\cite{zanella2023bayesian} or auditing in different privacy definitions~\cite{nasr2023tight}.

In general, the auditing algorithm $\mathcal{A}(\mathcal{T}|\delta, 1-\beta)$ takes the input as the implementation of training algorithm $\mathcal{T}$ and outputs the estimated privacy risk lower bound $\epsilon_L$ under a certain confidence $1-\beta$ given a fixed $\delta$.
We note that the output of $\mathcal{A}$ is also influenced by the input of $\mathcal{T}$ (the training data $D$ and the model $\theta_0$) and the its hyper-parameters such as $\eta$ or $B$ in \Cref{eq:dpsgd}.
The greatest privacy lower bound that $\mathcal{A}$ can estimate is $\epsilon_O$ when all guesses $\hat{S}$ are correct, which is limited by the statistical power given $T$ observations. For example, $\epsilon_O=5.6$ with $T=1,000$ trials for \Cref{eq:nasr_lb}.
For an effective DP auditing, we need $\epsilon_L \gtrsim \epsilon$.

\subsection{O(1) Framework Overview}\label{sec:o1_framework}
Instead of repeating training $\mathcal{T}$ and inference $f_\text{MIA}$ for $T$ pairs of $D\simeq D^\prime$ for obtaining $T$ observations, O(1) framework~\cite{steinke2023privacy} is efficient as it obtains $m\approx T$ observations in one run by excluding or including each of $m$ audit samples.
Given $n=|D|$, we have:
\begin{itemize}
    \item The auditor samples a membership encoding vector $S \in \{-1, 1\}^m$ with each element independently drawn from the $\texttt{Bernoulli}(\frac{1}{2})$ distribution, with $0\leq m \leq n$ and set $S_i=1$ for $i\in [n] \backslash [m]$\footnote{$[n]=\{0, 1, \cdots, n-1\}$}.
    \item The trainer trains $\theta = \mathcal{T}(\theta_0, D_\text{IN})$ with subset $D_\text{IN}=\{z_i|S_i=1, z_i\in D\}_{i=1}^m$ and outputs the sorted confidence $I=\{f_\text{SCORE}(z_i, \theta)\}_{i=1}^m$.
    \item The auditor makes membership predictions $\hat{S} = \{f_\text{MIA}(I_i) \}_{i=1}^m \in \{-1, 0, 1\}^m$ with $+1(-1)$ as the positive (negative) prediction and $0$ as abstention for unconfident guesses.
    \item The estimated risk lower bound is $\epsilon_L=f_\text{EST}(S, \hat{S})$.
\end{itemize}

In $f_\text{EST}$, the number of correct guesses $W:=\sum_i^m \{0, \hat{S}_i\cdot S_i\}$ can be estimated over $m$ guesses.
Suppose the null hypothesis is $H_0: \mathcal{T}$ is $(\epsilon, \delta)$-DP, the main result of O(1) in \Cref{thm:o1_main} indicates that if $H_0$ stands, $W \leq \frac{r\cdot e^\epsilon}{e^\epsilon+1} + O(\sqrt{r})$ with high probability.
Otherwise, we reject $H_0$ and believe $\mathcal{T}$ violates the claimed DP.
Then $\epsilon_L$ can be estimated by converting this hypothesis test into the confidence interval by finding the largest $\epsilon$ that we can reject at a desired confidence.
The optimal estimation $\epsilon_O$ in O(1) scales up with the number of observation $m$. \looseness=-1
 
\begin{theorem}[Main Result of \cite{steinke2023privacy}]\label{thm:o1_main}
    Let $(S, \hat{S}) \in \{-1,+1\}^m \times \{-1, 0, +1\}^m$ be the membership status and guess. 
    Assume the training algorithm $\mathcal{T}$ satisfies $(\varepsilon,\delta)$-DP.
    Let \( r := \left| \{ x \in \hat{S} \mid x \neq 0 \} \right| \) to be the number of guess, then for $v \in \mathbb{R}$,
    \begin{equation}
    \pr{}{\sum_i^m \max\{0, \hat{S}_i \cdot S_i\} \ge  v} \le \beta + \alpha\cdot m\cdot \delta, 
    \end{equation}
    where     
    \begin{align}
        \beta &= \pr{\check{W}^*}{\check{W} \ge  v}, \check{W^*} := \texttt{Binomial}(r, \frac{e^\epsilon}{e^\epsilon+1}) \\
        \alpha &= \max_{i\in[m]} \frac{2}{i} \pr{\check{W}^*}{v > \check{W} \ge  v-i}.
    \end{align}
\end{theorem}

Orthogonal to the O(1) framework, auditor can construct the audit dataset $D_{\text{audit}} = \{ \tilde{z}_i \}_{i=1}^m \subseteq D$ with each canary crafted from $\tilde{z}_i = f_\text{CRF}(\tilde{z}_i)$.
For example, a widely applied way is mislabeling the original sample with a random label~\cite{steinke2023privacy}.

\subsection{Threat Model}\label{sec:threat_model}
\subsubsection{O(1) Auditing Game}
In the auditing game, there are two parties: the \textit{trainer} curates the DP training implementation and runs $\mathcal{T}$ while the \textit{auditor} performs auditing algorithm $\mathcal{A}$.
The trainer claims that $\mathcal{T}$ satisfies $(\epsilon, \delta)$-DP, but is potential to violate the given DP if the implementation is flawed.
Therefore, the auditor aims to estimate the empirical risk lower bound $\epsilon_L$ for verifying the integrity of $\mathcal{T}$.
Formally, we summarize the whole auditing algorithm as:
$$
\mathcal{A}=f_\text{CRF}\circ \texttt{Bernoulli}(\frac{1}{2}) \circ \mathcal{T} \circ f_\text{SCORE} \circ f_\text{MIA} \circ f_\text{EST} ,\label{eq:aud_line}
$$
where the auditor performs all components except the training algorithm $\mathcal{T}$, and the auditor only query one run of $\mathcal{T}$.
Multiple training runs increase the number of empirical observations, which further improve the lower bound estimation.

\subsubsection{Limited Auditing Capability}
Before training, the auditor will send the dataset to the trainer with a designated model architecture.
Besides, we assume a strict and practical capabilities for auditors as follows:
\begin{itemize}
	\item \textbf{No manipulation on implementation}: the auditor does not interfere the DP training implementation, including gradient calculation or hyper-parameters. Existing works may modify the normal DP training step for maximizing auditing capability, such as by ignoring updates of non-audit samples or crafting the real gradients, while we do not assume such capability.
    \item \textbf{Normal training configuration}: we assume all configurations are normal without requirement of specific model initialization or hyper-parameter tuning (including learning rate or batch size). While existing works shown that using privacy-backdoor~\cite{liu2024precurious, wen2024privacy}, an extremely large learning rate or sub-sampling ratio helps to improve $\epsilon_L$, we do not force such assumption for a practical use.
    \item \textbf{Black-box access}: the auditor cannot obtain the target model's parameters and only audit through a black-box access. Existing works show that black-box is much more challenging than white-box in O(1) framework, and can only achieve $\epsilon_L\approx 1.3$~\cite{steinke2023privacy, panda2025privacy} for $\epsilon=4$ given confidence $\alpha=0.95$ and $m=1,000$.
    \item \textbf{Final model access}: the auditor can only get access to the final model in our setting, while existing works~\cite{nasr2021adversary} assume auditors can access all intermediate models.
\end{itemize} 

\section{No Free Lunch for O(1) Auditing}\label{sec:no_free_lunch}
Even though O(1) framework in \Cref{sec:o1_framework} is efficient, it sacrifices the auditing power due to two main challenges.
For illustrating the key challenges, we train CNN model on CIFAR10 dataset from scratch and perform a grid-search over hyper-parameters that may influence both auditing power and model utility as shown in \Cref{fig:demo_challenge}.

We consider two versions of canary function $f_\text{CRF}$ in O(1): 1) \textit{in-distribution} uses the original CIFAR10 subset as $D_\text{audit}$; 2) \textit{mislabeled} flips the true labels, and one variant with stronger $f_\text{MIA}$ 3) \textit{poisoned} applies privacy backdoor~\cite{liu2024precurious,wen2024privacy}.

\subsection{Inevitable Data Dependency}
While in O($T$) auditing, each observation is obtained from \textit{independent} training runs.
Observations obtained in O(1) are partially \textit{dependent},
as including or excluding one sample influences other audit samples.
And such data dependency influences in both training $\mathcal{T}$ and inferences $f_\text{SCORE}\circ f_\text{MIA}$.
For example, similar samples with the same labels in $D_\text{audit}$ are essentially correlated by the core feature mapping to the label.
Thus, these samples have a regularization effect~\cite{wu2020generalization} to each other, reducing the sample-specific memorization in $\mathcal{T}$.
Consequently in the inference phase $f_\text{SCORE}\circ f_\text{MIA}$, member scores are less distinguishable, resulting in weak membership inference attack (MIA).

\begin{figure}[thb]
    \centering
    
    \begin{subfigure}[t]{\linewidth} %
        \includegraphics[width=\columnwidth]{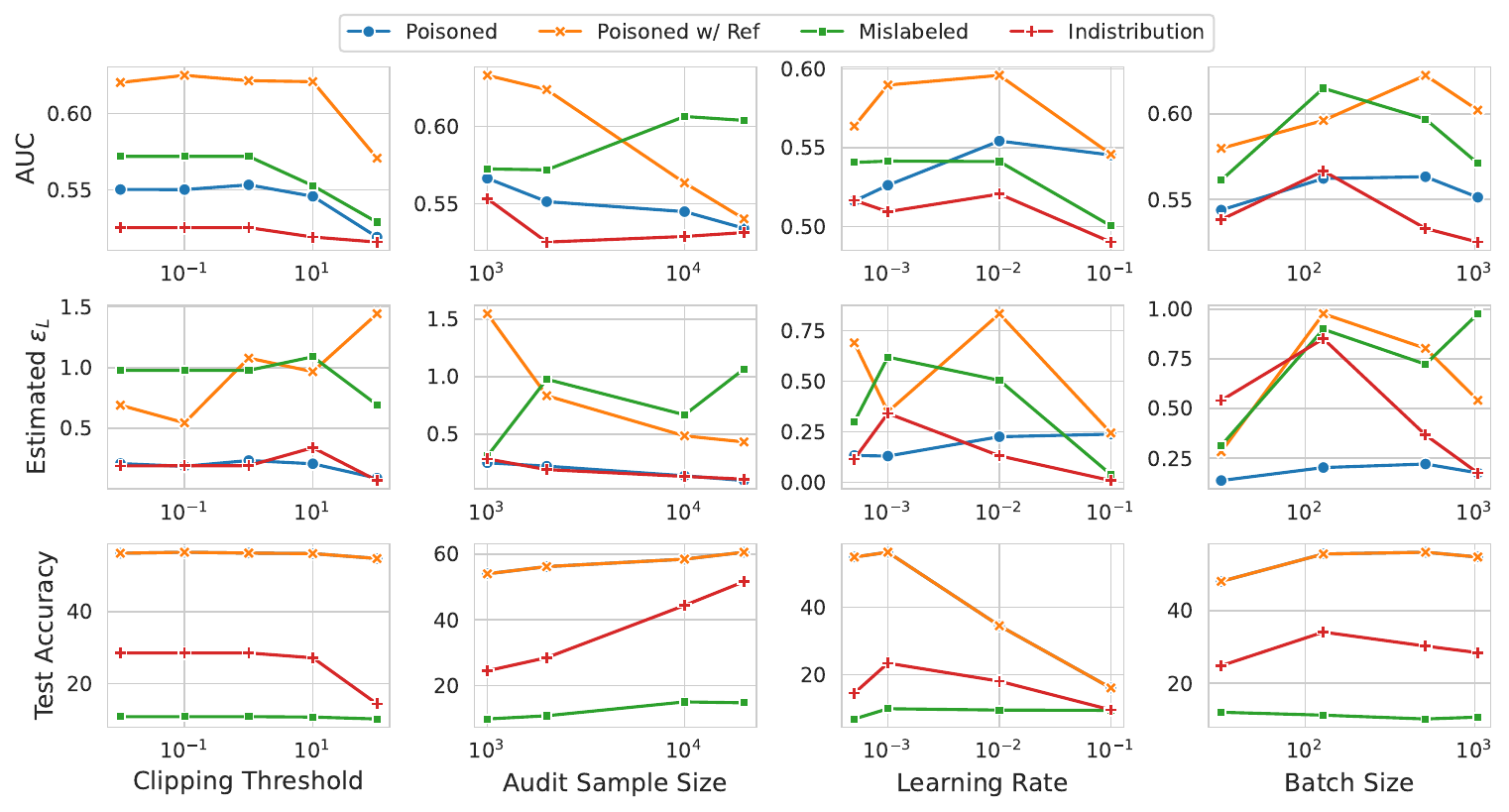}
        \caption{$\epsilon_L$ fluctuates with hyper-parameters, not scaling with $m$.}
        \label{fig:demo_challenge1}
    \end{subfigure}
    
    \vspace{0.5em} 
    
    \begin{subfigure}[t]{\linewidth} %
        \centering
        \includegraphics[width=0.48\columnwidth]{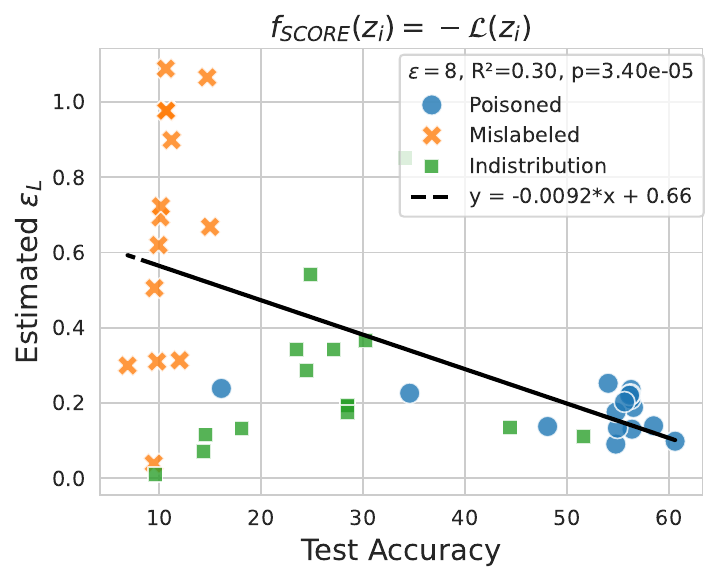}
        \hfill %
        \includegraphics[width=0.48\columnwidth]{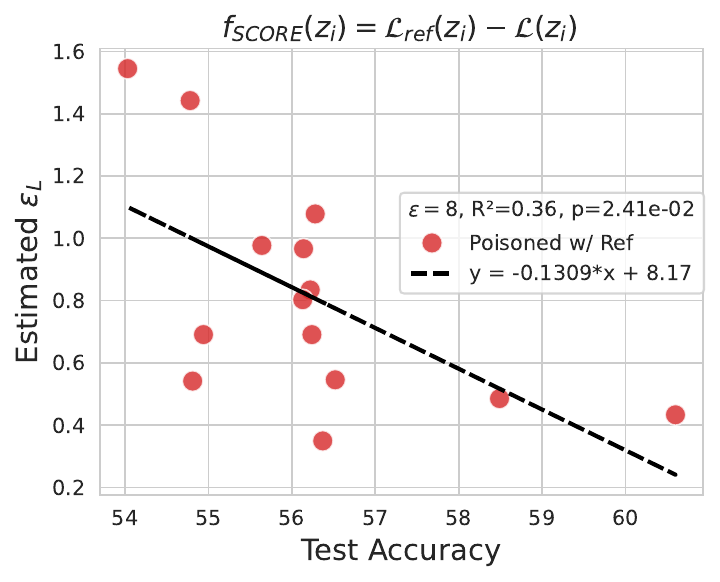}
        \caption{Trade-off between Utility and Auditing for O(1)}
        \label{fig:demo_challenge2}
    \end{subfigure}
    
    \caption{Preliminary analysis for O(1) Auditing.
    Fig (b) is summarized from checkpoints in Fig (a).
    By default, we use $m=2,000, B=2,000, \eta=1e-3, E=100$ and the theoretical epsilon is $\epsilon=8$.
    For \textit{Poisoned} baseline, the adversarial model initialization is warmed up on an extra auxiliary dataset with $5,000$ samples and 50 epochs.
    }
    \label{fig:demo_challenge}
\end{figure}

As the result shown in the second column of \Cref{fig:demo_challenge}(a), both the estimated $\epsilon_L$ and MIA AUC decrease with larger $m$ when audit canaries are real samples with correlation.
For mislabeled baseline with random labels (less dependency), $\epsilon_L$ scale up with $m$ while also fluctuates.
In summary, the inevitable data dependency limits auditing power of O(1) framework from two aspects:
\begin{itemize}
    \item Limited number of observation: the optimal estimation $\epsilon_O$ (when $f_\text{MIA}$ is 100\% accurate) is \textbf{not} positively correlated with $m$ as expected; thus, it is \textbf{impossible} to spot privacy violation when $\epsilon_O < \epsilon$, no matter how strong is $f_\text{MIA}$.
    \item Limited MIA capability: the dependent scores impair the MIA capability which directly result in weaker $\epsilon_L$.
\end{itemize}

\subsection{Implicit Conflict between Auditing and Utility}
We are the first to reveal that there is an implicit conflict between auditing and utility, which also limits the auditing power of O(1).
As shown in \Cref{fig:demo_challenge}(a), $\epsilon_L$ is sensitive to hyper-parameter choices.
While existing works mainly focus on tighter estimation and opt for a larger learning rate $\eta$ or batch size $B$ to maximize the auditing result, once the auditor attempts to balance the auditing-utility trade-off~\cite{panda2025privacy}, auditing power is clearly compensated as the negative trend shown in \Cref{fig:demo_challenge}(b).
Specifically, \textit{poisoned} seems has higher test accuracy and higher $\epsilon_L$ with advanced $f_\text{SCORE}\circ f_\text{MIA}$, it requires extra \textbf{non-private} auxiliary dataset, which is not comparable to \textit{mislabeled} and \textit{in-distribute}.
Without such assumption, \textit{mislabeled} has \textbf{no meaningful} utility as shown in \Cref{fig:demo_challenge}(b)-left while achieves higher $\epsilon_L$ than \textit{in-distribute}.

\subsection{Our Key Insights from the Generalization Perspective}\label{sec:insight}
We observe that the key to addressing the above two challenges lies in the boundary between sample-specific memorization and generalization.

\begin{figure}
    \centering
    \includegraphics[width=0.48\linewidth]{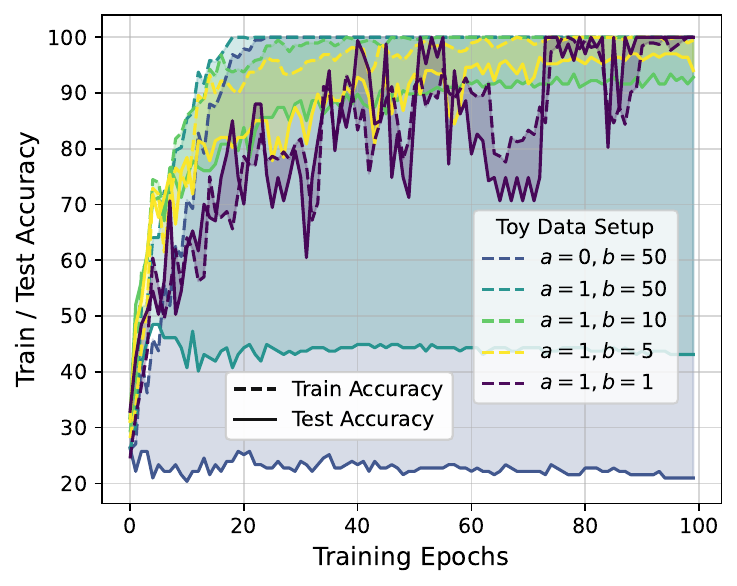}
    \includegraphics[width=0.48\linewidth]{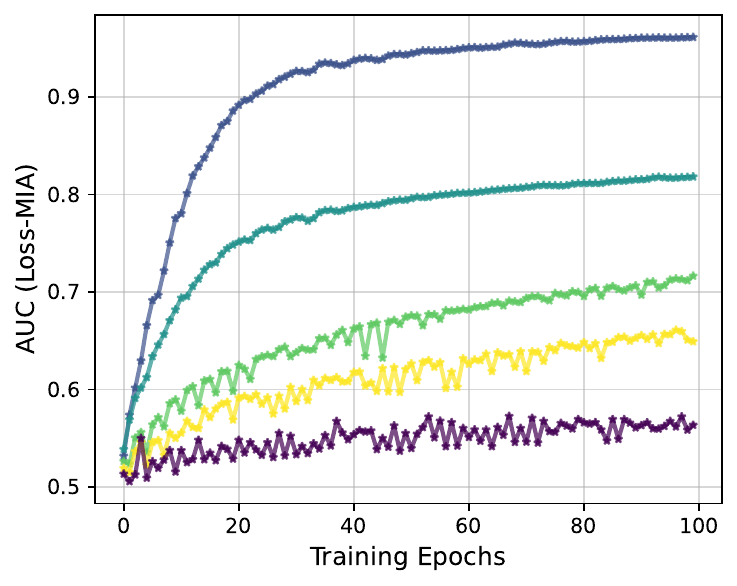}
    \caption{Interplay between sample-specific memorization (less data correlation for auditing) and generalization (more data correlation for utility).}
    \label{fig:demo_conflict}
\end{figure}

\textbf{Setup for Toy Experiments.}
To investigate the interplay, we construct a toy dataset that serves as an abstracted instance of real-world data, without loss of generality.
In our setup, each training sample ideally consists of informative features that determine its correct label.
Given a label $y$, we generate the corresponding sample as
\begin{equation}\label{eq:toy_data}
    \mathcal{X}_{y} = a\mathcal{N}(y, \sigma_0^2) + b\mathcal{N}(0, \sigma^2_0)
        \in \mathbb{R}^{d},
\end{equation}
where the first term represents the core feature that depends solely on the label $y$, and the second term adds sample-specific Gaussian noise with the basic noise scale $\sigma_0=0.1$.
We use $a \in \{0, 1\}$ to toggle the presence of label-dependent features, and $\alpha=1$ simulates real-data with correlated features and labels.
The second term $b \geq 0$ controls the level of sample-specific variation.
We optimize two-layer neural network with ReLU activations with SGD on the toy data for 100 epochs.

\noindent\textbf{Key Insight I. Uncorrelated Data for Less Dependency.}
Fixing the noise level $b=50$, \Cref{fig:demo_conflict} shows that $a=0$ with correlated labels leads to larger generalization gap (left) and higher MIA AUC (right) than $a=1$ which is more like real-world data. 
Fixing the correlation level $a=1$, larger relative sample-specific noise level $b/a$ enhances MIA.
This reveals the fact that auditing power depends on the sample-specific noise and the correlation level between feature and label impairs the relative sample-specific signal.
Therefore, we are motivated to design $f_\text{CRF}$ by constructing $D_\text{audit}$ with uncorrelated data for reducing  dependency in $\mathcal{T}\circ f_\text{SCORE}\circ f_\text{MIA}$.

\noindent\textbf{Key Insight II. Decoupling Auditing from Utility.}
Real-data definitely has correlated feature and labels for meaningful test performance.
Given $a=1$ in \Cref{fig:demo_conflict}, the generalization gap decreases and the testing accuracy gets better with smaller sample-specific noise level $b$, while the MIA AUC gets worse.
The implicit conflict is clear, as the auditing power relies on the sample-specific noise, which hurts the model generalization capability for good utility.
This tension motivates us to challenge the necessity of coupling utility and auditing in O(1).
Our intuition is two fold:
\begin{itemize}
\item \textit{Decoupled Criterion.} As mentioned in \Cref{sec:dpsgd}, the definition of DP rely on nothing but $\mathcal{T}$. 
Therefore, as we summarized in \Cref{tab:goal}, meaningful model utility for auditors is not a necessary criterion for \textbf{Case I} where the goal is to diagnose DP implementation with $\epsilon_L=\mathcal{A}(\mathcal{T}|*)$.
The auditing-utility trade-off should be considered as criterion in \textbf{Case II} where $\epsilon_L=\mathcal{A}(\mathcal{T}|D, \theta)$ is expected to be conditioned on a specific run.
\item \textit{Decoupled Objective.} 
For \textbf{Case II} with coupled criterion, our idea is to separate training objectives for utility and training for auditing.
The utility is optimized with main task objective, while we formulate auditing goal as a membership encoding problem.
\end{itemize}

Based on our key insights, we will improve the O(1) framework with a data-independent framework \ours~ for Case I in \Cref{sec:audit_only}, and a data-dependent O(1) framework \ourss~ for Case II in \Cref{sec:audit_utility}.

\begin{table}[t]
\centering
\caption{Overview of O(1) auditing variants without manipulation training algorithm and black-box access to the final model.
(\yes~ indicates `Yes', \no~ indicates `No', $\emptyset$ means $D_\text{audit}=D$ and $n=m$)
}\label{tab:goal}
\resizebox{0.5\textwidth}{!}{
\begin{tabular}{c|cc|c|cc}
\toprule[1pt]
\multirow{2}{*}{\begin{tabular}[c]{@{}c@{}}O(1) Audit Variants\end{tabular}} 
    & \multicolumn{2}{c|}{Data Correlation} 
    & \multirow{2}{*}{\begin{tabular}[c]{@{}c@{}}Need \\ $D_\text{aux}$?\end{tabular}}
    & \multicolumn{2}{c}{Criterion} \\
\cline{2-3} \cline{5-6}
    & \multicolumn{1}{c|}{$D_\text{audit}$} & $D \setminus D_\text{audit}$ 
    &                                      & Utility & Auditing \\
\midrule
In-distribution                                                                                                                   & \multicolumn{1}{c|}{\yes}      & \yes          & \no                      & \yes             & \no                               \\ \cline{2-6}
Mislabeled Data                                                                                                                   & \multicolumn{1}{c|}{\no}       & \yes           & \no                      & \yes             & \yes                              \\ \cline{2-6}
Poisoned Model                                                                                                                    & \multicolumn{1}{c|}{\yes }      & \yes           & \yes                      & \yes             & \yes                              \\ \midrule[0.8pt]
\ours~ (Data-Independent)          & \multicolumn{1}{c|}{\no}       & $\emptyset$          & \no                      & \no            & \yes              \\ \cline{2-6}
\ourss~ (Data-Dependent)            & \multicolumn{1}{c|}{\no}       & \yes          & \no                      & \yes            & \yes                \\ \bottomrule[1pt]
\end{tabular}
}
\end{table}

\section{Data-Independent O(1) Auditing Framework}\label{sec:audit_only}
In this section, we propose data-independent framework \ours~, which aims to improve auditing power within O(1) complexity by reducing data dependency via carefully designed $f_\text{CNR}, f_\text{SCORE}$ and $f_\text{MIA}$.

\subsection{Building Canaries for Membership Encoding}
We first resolve data dependency problem with better $f_\text{CNR}$.

\subsubsection{Rules of Thumb for Audit Canaries}
As summarized in \Cref{tab:canary}, many canary strategies used in $O(T)$ auditing~\cite{jagielski2020auditing, nasr2023tight} are not applicable to $O(1)$.
For designing the synthetic audit data, here are rules of Thumb for canaries in O(1) framework:
\textbf{a) Independence:} Each canary should contribute independently to the audit, unlike methods such as PGD~\cite{nasr2023tight}, Orthogonal~\cite{nasr2023tight}, or ClipBKD~\cite{jagielski2020auditing}, which assume $O(n)$ settings.
\textbf{b) Easy-to-Spot:} Canaries should be easily distinguishable to enable tighter auditing bounds.
\textbf{c) Scalable Generation:} $O(1)$ auditing benefits from many canaries in a single run, making per-sample optimization or reliance on auxiliary data impractical.
\begin{table}[t]
\centering
\caption{Canaries in DP auditing (\yes indicates `Yes')}
\label{tab:canary}
\resizebox{0.5\textwidth}{!}{
\begin{tabular}{c|ccc|cc}
\toprule[1pt]
Canary  & Real-data? & Need $D_\text{aux}$? & Optimization? & For O(1)? & For black-box? \\
\hline
Blank          & \no              & \no                & \no           & \no        & \yes            \\
\hline
In-distribution & \yes             & \no                & \no           & \yes       & \yes            \\
Mislabel       & \yes             & \no                & \no           & \yes       & \yes            \\
\hline
PGD            & \yes             & \no                & \yes          & \no       & \no             \\
Orthogonal     & \yes             & \yes               & \yes          & \no       & \no             \\
ClipBKD        & \yes             & \yes                & \yes           & \no       & \no            \\
\hline
Ours           & \no              & \no                & \no           & \yes       & \yes       \\  
\bottomrule[1pt]
\end{tabular}
}
\end{table}

\subsubsection{Uncorrelated Pair-Matching as Membership Encoding}
Based on \Cref{sec:insight}, the essence of auditing is better membership encoding, with the goal to minimize loss of member samples while maximize the loss of non-member samples.
Unlike prior work in non-DP settings, our auditor cannot modify the training procedure; thus the key is constructing the joint space of audit data for better membership encoding.

We propose to use \textbf{uncorrelated} pair-matching task for membership encoding, with the objective to minimize the probability of matching a random feature to a random label:
\begin{equation}
\mathcal{L}_\text{ME} = \frac{1}{m}\sum_{i=1}^m
\mathbb{I}(S_i=1)\mathcal{L}_\text{CE}(f_\theta(x_i), y_i), \label{eq:loss_me}
\end{equation}
where $x_i$ and $y_i$ are drawn independent from $\mathcal{X}$ and $\mathcal{Y}$.
Without loss of generality, we instantiate as a multi-label classification problem with $\mathcal{X}\subset \mathbb{R}^d$ and $\mathcal{Y}\in [C]$ of $C$ classes, leaving auto-regressive extensions in \Cref{sec:lm_extend}. 

We leverage this high-dimensional sparsity by independently sampling \(m\) audit points uniformly from all possible input–label combinations from $\mathcal{X}\times \mathcal{Y}$, indicating that each $(x_i, y_i)\in D_\text{audit}$ is very likely separated from others.
As \(|\mathcal X|\) and \(|\mathcal Y|\) grow, the joint space expands exponentially, making it virtually impossible for the model to generalize as for unseen non-member samples $(x_i, y_i) \in D_\text{audit}$ with $S_i=-1$:
\[
\mathbb{E}_{(x,y)\sim \mathcal{X}\times\mathcal{Y}}[p_\theta(y|x)] \approx \frac{1}{C} \implies \mathcal{L}_\text{CE}(x_i,y_i) \approx \log C
\]
creating an inherent loss gap between members (\(\mathcal{L}_\text{CE} \to 0\)) and non-members (\(\mathcal{L}_\text{CE} \approx \log C\)).

\subsubsection{Synthetic Canary Generation}
Now we discuss concrete construction for $\mathcal{X}$ and $\mathcal{Y}$.
We notice that pure random $\mathcal{X}$ with uncorrelated labels $\mathcal{Y}$ is the ideal solution that satisfies above three rules and fully expand the joint space.
Especially when $\mathcal{X}$ is pure random (as $a=0$ in \Cref{fig:demo_conflict}), samples are more separated than natural images which center around true labels, leading to faster convergence rate~\cite{zhang2016understanding} than only random label.
We propose synthetic canary generation with two variants in \Cref{alg:canary}.
Specifically for orthogonal features, output features are approximately orthogonal when $n>d$.
\begin{algorithm}\label{alg:canary}
\caption{Synthetic Data Generation for $D_\text{audit} \in \mathcal{X}\times\mathcal{Y}$}
\begin{algorithmic}[1]
\Require Number of audit samples $m$; $\mathcal{X}$ with mode $\in$ \{gaussian, orthogonal\}, feature dimensions $d$, and sample-noise scalar $\sigma_0$; $\mathcal{Y}=\texttt{Uniform}([C])$ with $C$ classes
\Ensure Synthetic dataset $D_\text{audit}=\{(x_i, y_i)\}_{i=1}^m$
\If{mode == \texttt{orthogonal}}
    \State \commentblue{Generate orthonormal matrix $Q$}
    \State $Q, R \gets \text{QR}(\mathcal{N}(0,1)^{d \times d})$
    \State \commentblue{Normalize coefficient vectors}
    \State $U \gets \mathcal{N}(0,1)^{m \times d}$
    \State $U \gets U / \|U\|$ row-wise
    \State $\{x_i\}_{i=1}^m \gets U \cdot Q^\top$
\ElsIf{mode == \texttt{gaussian}}
    \State $\{x_i\}_{i=1}^m \gets \mathcal{N}(0,\sigma_0^2)^{n \times d} \cdot \sigma_0$
\EndIf

\State $\{y_i\}_{i=1}^m=\{y_i|y_i\in \mathcal{Y}, i\in [m]\}$
\State \Return $D_\text{audit}$
\end{algorithmic}
\end{algorithm}

\subsubsection{Model Architecture}
By default for classification task, we use a 2-layer neural network with ReLU activations.
Following previous work~\cite{zhang2016understanding}, a two-layer ReLU neural network with $2n + d$ parameters can universally represent any function defined on a sample of size $n$ in $d$-dimensional space.
In general, the synthetic $D_\text{audit}$ is compatible with different model architectures with input dimension $d$ and output size $C$.

\subsection{Inclusive auditing by self-comparison}
\newcommand{\tin}{{\text{IN}}}
\newcommand{\tout}{{\text{OUT}}}
\newcommand{\xin}{x_{\text{IN}}}
\newcommand{\xdata}{x_{\text{FIXED}}}
\newcommand{\xout}{x_{\text{OUT}}}
\newcommand{\nin}{{n_{\text{IN}}}}
\newcommand{\nout}{{n_{\text{OUT}}}}
\newcommand{\scin}{s_{\text{IN}}}
\newcommand{\sout}{s_{\text{OUT}}}
\begin{algorithm}[h]
    \caption{Data-Independent O(1) Auditing (synthetic canary + self comparison)}\label{alg:comp}
    \begin{algorithmic}[1]
        \State \textbf{Input:} training algorithm $\mathcal{T}$, canary distribution $\mathcal{X}\times\mathcal{Y}$
        , audit confidence level $1-\beta=0.95$
        \State \textbf{Output:} estimated privacy lower bound $\epsilon_L$
        \State Auditor gets audit set $D_\text{audit} \gets f_\text{CRN}(m, \mathcal{X}\times\mathcal{Y})$ and comparison set as $D_\text{comp}\gets\{x_i, y^\prime|y^\prime\in \mathcal{Y}\}_{i=1}^m$
        \State Trainer trains on audit data $\theta = \mathcal{T}(D_\text{audit}|\theta_0)$
        \State Sample $m$ evaluation samples $D_\text{mia}$, with membership status $S_i\in \{-1,+1\}$ drawn from \texttt{Bernoulli($\frac{1}{2}$)}:
        \State \quad \quad $z_i \gets \left\{
            \begin{array}{ll}
            D_\text{comp}[i]; \tilde{z}_i=D_\text{audit}[i]  & S_i=-1 \\
            D_\text{audit}[i]; \tilde{z}_i=D_\text{comp}[i] & S_i=1
            \end{array}
            \right.$
        \State Compute scores for $D_\text{mia}$ as $I=\{f_\text{SCORE}(z_i, \theta)\}_{i=1}^m$ where $f_\text{SCORE}(z_i)=\mathcal{L}(f_\theta, \tilde{z}_i)-\mathcal{L}(f_\theta, z_i)$
        \State $f_\text{MIA}$ sorts $I$ and make $r$ guesses with $r/2$ samples with highest (w.r.t. lowest) $I_i$ as $\hat{S}=1$ (w.r.t. $\hat{S}=-1$)
        \State Estimate privacy risk lower bound $\epsilon_L$=$f_\text{EST}(S, \hat{S}, \delta, 1-\beta)$
    \end{algorithmic}
\end{algorithm}

Uncorrelated synthetic canary constructed to enlarge the generalization gap with the pair-matching task makes scores less dependent and more distinguishable.
Now we further improve MIA capability with better $f_\text{SCORE} \circ f_\text{MIA}$.
The key intuition that we can further improve MIA for auditing is attributed to the label independency~\Cref{property:y}.
\begin{property}\label{property:y}
\textit{Label Independency:} For all $x \in \mathcal{X}$ and $y, y' \in \mathcal{Y}$, the function $f_\text{CRN}$ in \Cref{alg:canary} satisfies
\[
\Pr_{\mathcal{X} \times \mathcal{Y}}[y \mid x] = \Pr_{\mathcal{X} \times \mathcal{Y}}[y' \mid x].
\]
\end{property}

Existing works~\cite{steinke2023privacy} use $f_\text{SCORE}=-\mathcal{L}(f_\theta(x_i), y_i)$ and then sort scores for $f_\text{MIA}$ given negative and positive thresholds, resulting in guesses dependent on sample-hardness of $D_\text{audit}$.
This issue is unique for O(1) because each observation in O(T) is only obtained by differing one sample.
And as the original work points out, it is non-trivial to leverage advanced MIAs with reference models to calibrate each sample hardness.

Instead, we propose \Cref{alg:comp} which makes $f_\text{MIA}$ independent among audit samples by calibrating the score $f_\text{SCORE}$ with a counterfactual calibration.
Specifically, we make the following modifications compared to the original O(1) framework:
a) We train on the full set of $D_\text{audit}$ and construct non-member set by  independently sample a fresh label from $\mathcal{Y}$.
b) We independently flip the observation target via $\texttt{Bernoulii}(\frac{1}{2})$, ensuring an unbiased $f_\text{MIA}$.
c) We only need one-side threshold because the counterfactual calibrated $f_\text{SCORE}$ is symmetric.
Thus, we have \Cref{thm:enhanced} with the key prerequisite of \Cref{property:y}, with proof in Appendix.

\begin{theorem}\label{thm:enhanced}
    Assume training algorithm $\mathcal{T}$ satisfies $(\epsilon,\delta)$-DP, and $f_\text{CNR}$ in \Cref{alg:canary} is used in auditing \Cref{alg:comp}, 
    then the inequality in \Cref{thm:o1_main} hold for $(S, \hat{S}) \in \{-1,+1\}^m \times \{-1,0,+1\}^m$ in \Cref{alg:comp}.
\end{theorem}

\section{Data-dependent O(1) Auditing Framework}\label{sec:audit_utility}
In this section, we propose data-dependent framework \ourss~ with aims to improve the auditing-utility trade-off given good efficiency with O(1) complexity for Case II.
This scenario is practical when auditor aims to estimate the conditioned empirical risk for a specific DP training run given data $D$ and model $\theta$, or when one DP training is expensive thus $\theta$ is required to have meaningful utility.

\subsection{Multi-Task for Decoupled Objective}\label{sec:decouple_obj}
While the conflict between utility and auditing is inherent as shown in \Cref{sec:insight}, we note that it is because the in typical auditing training, the two training objectives are disentangled.
Specifically, $m$ original or mislabeled canaries still share the same target space $\mathcal{Y}$ with the rest $n-m$ non-audit samples in the training set $D$.

Our key insight is to decouple the objective of auditing and utility during DP training by augmenting training dataset with an extra feature and label space.
Thus, we transfer the data distribution as $\mathcal{X}\times\mathcal{G} \to \mathcal{Y}\times \mathcal{E}$, where $\mathcal{G}$ and $\mathcal{E}$ denote the trigger space and tag space for membership encoding.
Thus, combining the membership encoding objective in \Cref{eq:loss_me} with the typical main task, we have the training objective as
\begin{align}
    \mathcal{L} = \underbrace{\mathcal{L}_\text{CE}(f_\theta(x_i), y_i)}_{\mathcal{L}_\text{main}} 
    + \lambda \underbrace{\mathcal{L}_\text{CE}(f_\phi(x_i), e_i) \mathbb{I}(S_i=1)}_{\mathcal{L}_\text{ME}}, \label{eq:loss_multi}
\end{align}
where $\lambda$ is the coefficient for balancing encoding strength.
Here, $\phi$ denotes an additional linear head on top of the shared encoder, following the standard multi-task setup with separate heads for distinct objectives. This allows joint optimization of the main task and membership encoding.

For optimizing objective, we construct dataset in \Cref{alg:trigger_tag}, which returns $D_\text{multi}$ for calculating \Cref{eq:loss_multi}.
The self-comparison framework \Cref{alg:comp} is seamlessly integrated with the returned $D_\text{audit}$ with $D_\text{comp}$ by replacing notation of $\mathcal{X}\times\mathcal{Y}$ with $\mathcal{G}\times\mathcal{E}$ and $f_\text{SCORE}$ is calculated on the separated encoding loss $\mathcal{L}_\text{ME}$.

\begin{algorithm}\label{alg:trigger_tag}
\caption{Multi-Task Dataset Construction for \ourss~}
\label{alg:multi_task_data}
\begin{algorithmic}[1]
\Require Real dataset $D = \{(x_i, y_i)\}_{i=1}^n$; number of audit samples $m$; trigger generator $\mathcal{G}$; tag generator $\mathcal{E} = \texttt{Uniform}([C_e])$
\Ensure $\mathcal{D}_{\text{multi}}$ for training, $D_\text{audit}, D_\text{comp}$ for auditing
\State Sample $m$ indices $\mathcal{I}_\text{audit} \subset [n]$ uniformly at random
\State Initiate empty dataset $D_\text{multi}, D_\text{audit}, D_\text{comp}$
\For{$i = 1$ to $n$}
    \If{$i \in \mathcal{I}_\text{audit}$}
        \State Generate trigger $g_i \sim \mathcal{G}$ and sample tag $e_i \sim \mathcal{E}$ 
        \State Generate a fresh tag $e_i^\prime \sim \mathcal{E}$
        \State Apply trigger $x_i \gets x_i \circ g_i$ and set $S_i \gets 1$
        \State Add $(x_i, y_i, e_i, S_i)$ to $D_\text{audit}$ 
        \State Add $(x_i, y_i, e_i^\prime, S_i)$ to $D_\text{comp}$
    \Else
        \State $e_i \gets \texttt{None}$, $S_i \gets 0$
    \EndIf
    \State Add $(x_i, y_i, e_i, S_i)$ to $\mathcal{D}_{\text{multi}}$
\EndFor
\end{algorithmic}
\end{algorithm}

\subsection{Multi-Bit Membership Encoding}
Essentially, the joint space $\mathcal{G}\times\mathcal{E}$ induces sample-specific patterns through unique $\langle g_i, e_i \rangle$ pairs, enabling precise membership encoding. 
If two audit samples have same trigger-tag pair, they are not independent enough.
According to the birthday attack~\cite{bellare2001introduction}, the collision probability that two samples have same $\langle g_i, e_i \rangle$ is $C(|\mathcal{E}|, m) \approx \frac{m^2}{2|\mathcal{E}|}$ for $1\leq m \leq \sqrt{2|\mathcal{E}|}$.
Thus, the ideal number of audit samples for independent membership encoding is $m \ll \sqrt{2|\mathcal{E}|}$.

For a linear projection head mapping $d_h$-dimensional features to $|\mathcal{E}|$ classes, this requires $d_h \times |\mathcal{E}|$ parameters---a potential bottleneck for utility and efficiency of DP training at scale.
We propose multi-bit encoding which encodes each trigger with multiple tags $\{e_i \in \mathcal{E}\}_{i=1}^H$, thus expanding the number of unique tags to by a factor of ${\mathcal{E}}\choose{H}$.
Suppose $|\mathcal{G}|\cdot|\mathcal{E}|$ trigger tag combinations can cover $m$ sample-specific encoding, it requires $|\mathcal{E}|\cdot d_h$ number of weights in its linear head.
We can reduce the number of weights required for $m$ samples by a factor of $1/{{\mathcal{E}}\choose{H}}$, which achieves its minimum when $H=|\mathcal{E}|/2$ given even $|\mathcal{E}|$ is even; and $(|\mathcal{E}|+1)/2$ and $(|\mathcal{E}-1|)/2$ for odd $|\mathcal{E}|$.

\section{Extending to language modeling}\label{sec:lm_extend}
Our framework \ours~ and \ourss~ are general and scalable to other tasks besides image classification.
Now we exemplify that both proposed frameworks can be seamlessly extended to language models (LM).

For LMs in \ours~, we adapt the trigger-tag pair as a prefix-suffix pair matching problem.
Formally, in the auto-regressive task of LM, we denote the text sequence as $\mathbf{x}\in\mathcal{X}$ with prefix $\mathbf{x}_p \circ \mathbf{x}_s$.
We propose to leverage LM's capability on learning the repetition pattern to maximize the membership encoding, which is different from a recent work which uses random tokens~\cite{panda2025privacy}.
To avoid influence of existing tokens as~\cite{panda2024new}, we augment the token vocabulary from $V$ with $V_\text{new}$.
For each $\mathbf{x}_i\in D_\text{audit}$ for $i\in[m]$, we independently sample only two random tokens from the new token set $V_\text{new}$, and repeat them with $|\mathbf{x}_p|$ and $|\mathbf{x}_s|$ times to create one canary.
A freshly sampled token is sampled from the same $V_\text{new}$ for $\mathbf{x}\in D_\text{comp}$ by fixing the $t_p$-length prefix.

Then we can easily generalize above solution by adding $n-m>0$ samples of normal text with $S_i=0$ in \ourss~ for meaningful utility.
The role of token embeddings $V_\text{new}$ in our LM variant is equal to the extra task head for optimizing the second term in \Cref{eq:loss_multi} for classification.
One difference is that there is no need to merge such pair pattern in $n-m$ normal samples because the optimization over new tokens in $V_\text{new}$ and normal text tokens in $V$ are naturally separated.

\section{Experiments}
As the proposed framework is unified to different inputs, we evaluate \ours~ in \textbf{Case I} and \ourss~ in \textbf{Case II} across image classification tasks and language modeling tasks.

\begin{figure*}[h]
    \centering
    \begin{subfigure}[b]{0.32\textwidth}
        \includegraphics[width=\textwidth]{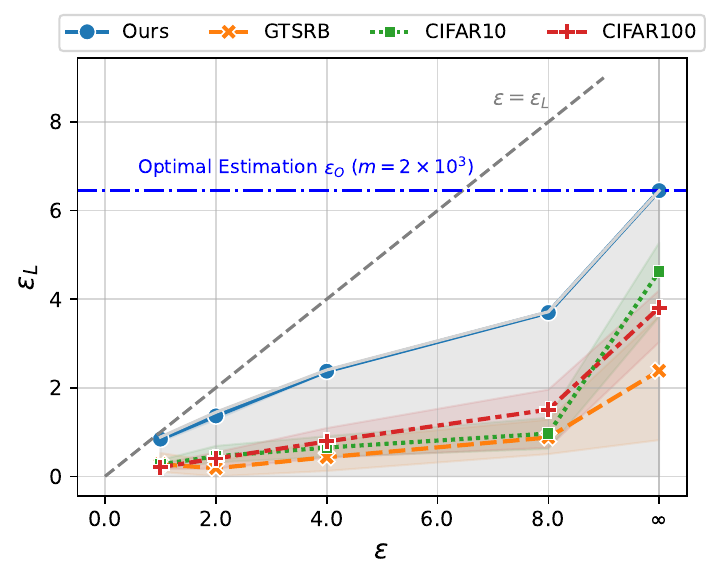}
        \caption{MLP}
        \label{fig:main_inf_eps_mlp}
    \end{subfigure}
    \hfill
    \begin{subfigure}[b]{0.32\textwidth}
        \includegraphics[width=\textwidth]{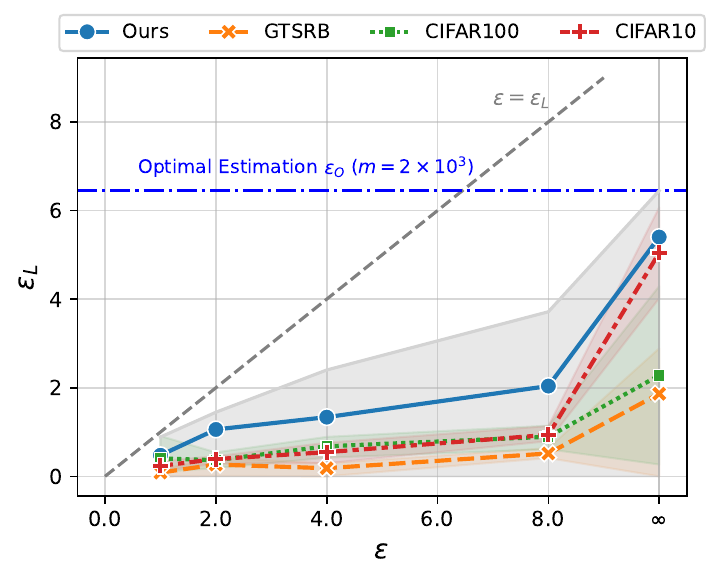}
        \caption{CNN}
        \label{fig:main_inf_eps_cnn}
    \end{subfigure}
    \hfill
    \begin{subfigure}[b]{0.32\textwidth}
        \includegraphics[width=\textwidth]{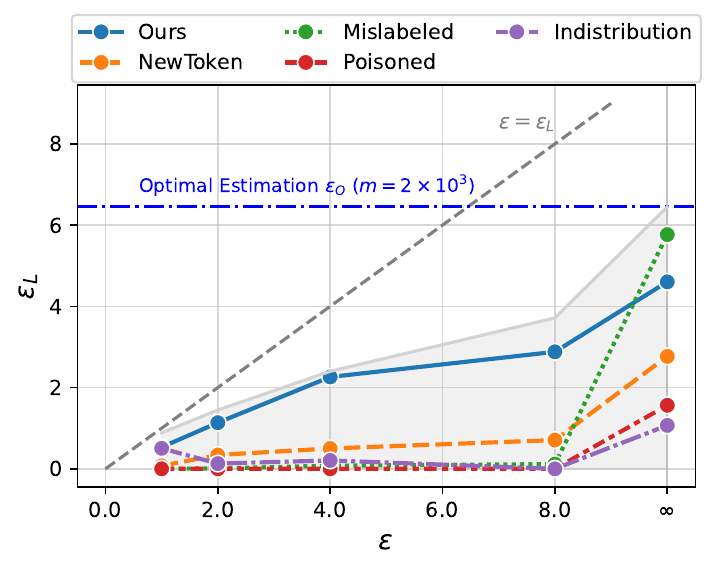}
        \caption{GPT-2 on PubMed}
        \label{fig:main_inf_eps_pubmed}
    \end{subfigure}
    \caption{Auditing improvement in Case I of data-independent setting with $\delta=1e-5$ and  95\% confidence.
    Gray shadows indicate the tightest $\epsilon_L$ achieved by the default MLP with orthogonal synthetic canary in \ours~ across models, which outperforms the state-of-the-art black-box performance~\cite{steinke2023privacy, panda2025privacy} (e.g., $\epsilon_L\approx 1.3$ for $\epsilon=4$), and approaches to white-box auditing performance (e.g., $\epsilon_L \approx 2.2$ for $\epsilon=4$) but saves at least $T=10^3$ training runs.
    In (a)(b), the range covered by each dataset (GTSRB, CIFAR10, and CIFAR100) summarizes $\epsilon_L$ across baselines (In-distribution, Mislabeled, and Poisoned).
    }
    \label{fig:main_inf_eps}
\end{figure*}

\subsection{Experimental Setups}
\textbf{Tasks and Dataset.}
In image classification task, for covering general types of model architectures, we evaluate convolutional neural network~\cite{papernot2021tempered} and transformer-based model of ViT~\footnote{\url{https://github.com/huggingface/pytorch-image-models/blob/main/timm/models/vision\_transformer.py}}, both of which are widely implemented in standard DP training benchmarks~\cite{papernot2021tempered, dormann2021not, bu2022automatic, panda2024new, de2022unlocking}.
To investigate the influence of model initialization, we consider both training from scratch, and fully fine-tuning over pre-trained parameters.
By default, the ViT encoder~\cite{dosovitskiy2020image} is pre-trained on ImageNet-21k, thus it is capable to capture general vision representation for classification.
Besides our synthetic audit data, we evaluate on standard CIFAR110 dataset used in existing auditing benchmark and additionally include the more difficult CIFAR100.
We also evaluate on GTSRB, which contains real-world traffic sign images with high inter-class similarity and varying image quality.\looseness=-1

In language modeling (LM) task, we follow previous work~\cite{panda2025privacy} and use GPT-2 family as an representative model architecture.
It should be noted that the improvement of auditing framework does not depend on specific architecture.
And following \cite{panda2025privacy}, we evaluate on PersonaChat dataset, and additionally include a subset of PubMed data in 2023, after GPT-2 pre-training data's cut off date in 2019, both datasets are widely used in previous DP training works~\cite{li2021large, yu2021differentially}.

\textbf{Baseline.}
In general, the key difference between data-independent (Case I) and data-dependent (Case II) setting is that $m=n$ in \text{Case I} for less influence from non-audit data for stronger auditing, while $m<n$ in Case II for maintaining acceptable utility.
For both cases, we compare with the following black-box O(1) variants with either different canary types and advanced MIA components.
\begin{enumerate}
	\item \textit{In-distribution} is the standard baseline of O(1) framework~\cite{steinke2023privacy} under black-box setting with a random subset of samples in the original training dataset as audit data. 
	\item \textit{Mislabel} represents a popular canary construction as also used in O(1)~\cite{steinke2023privacy} which keeps the original features but randomly flips the labels of the $m$ audit samples. As for LM tasks, it equals to a recent work~\cite{panda2025privacy} that constructs canary with in-distribution prefix and random tokens as suffix.
	\item \textit{Poisoned} is adapted from advanced MIA~\cite{liu2024precurious, wen2024privacy} by crafting the model initialization for amplifying the privacy risk of the training dataset, which represents a stronger auditing capability as it leverages extra assumptions beyond the auditing capability as we defined in \Cref{sec:threat_model}. We build the poisoned momel by warming up the model on an in-distribution auxiliary subset.
    \item \textit{NewToken}~\cite{panda2023differentially} is a state-of-the-art canary construction method specifically for LMs, by using new tokens as suffix. 
    We follow the best performed setting~\cite{panda2023differentially} by using random prefix and supervised fine-tuning (SFT) loss objective for DP auditing.
\end{enumerate}

\textit{Ours} indicates \ours~ in case I and \ourss~ in case II. 
For case I, we by default use a 2-layer MLP with ReLU network~\cite{zhang2016understanding} for image classification and use GPT-2 for language modeling task, because utility is not a necessary criterion.
By default, we set the input feature dimension as $d_x=10^3$, hidden states dimension as $d_h=10^5$ and output space size $C=10^3$ for maximizing the joint space size.

For image tasks in case II, \ourss~ is built on \ours~ by decoupling the training objective and revising the way of integrating sample-specific features in training data for maintaining utility. While for LM, the training objective and way of inserting canary feature keeps the same as we find the utility degradation is low especially for $n>m$ in case II due to the naturally separate space of normal and new tokens.

By default, we report results with $\delta=1\times10^{-5}$ and confidence 95\%.
We use the same training hyper-parameters for one type of model architectures, and we follow previous work to set the sub-sampling ratio for all DP training as 0.1.

\begin{figure*}[h]
    \centering
    \begin{subfigure}[b]{0.6\textwidth}
        \includegraphics[width=\textwidth]{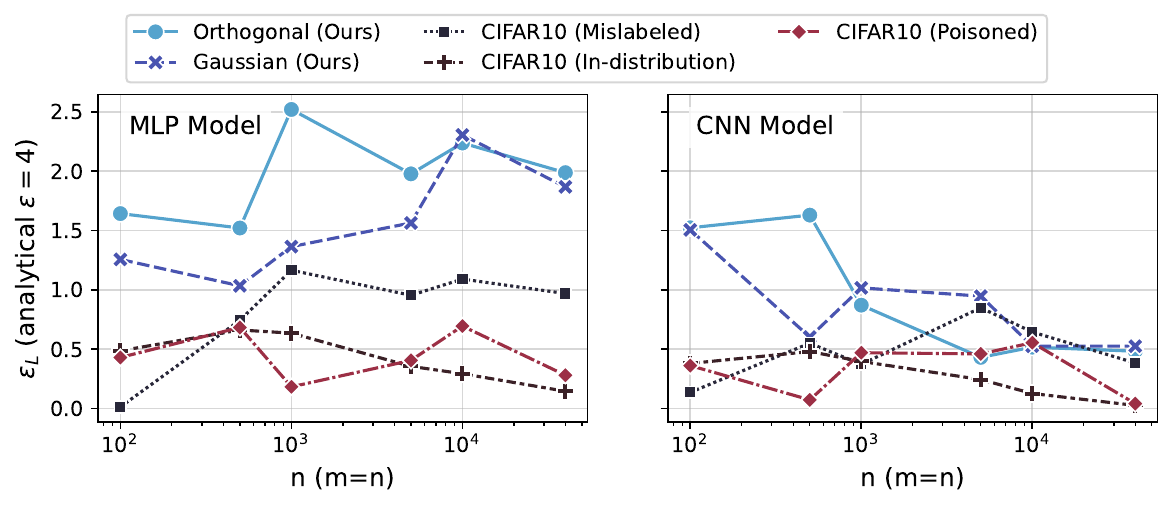}
    \end{subfigure}
    \begin{subfigure}[b]{0.34\textwidth}
        \includegraphics[width=\textwidth]{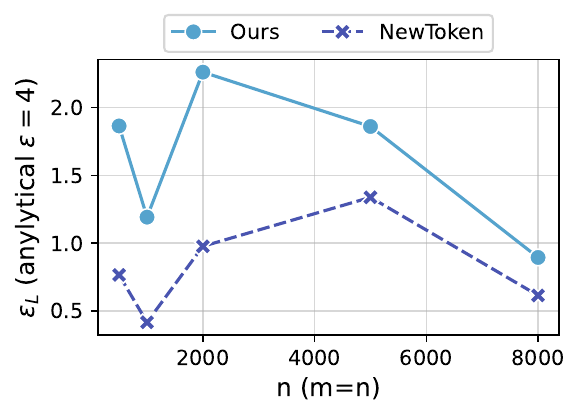}
    \end{subfigure}
    \caption{Influence of $m$ for image classification (left and middle) and language modeling (right)}
    \label{fig:inf_m}
\end{figure*}

\begin{figure}
	\includegraphics[width=\linewidth]{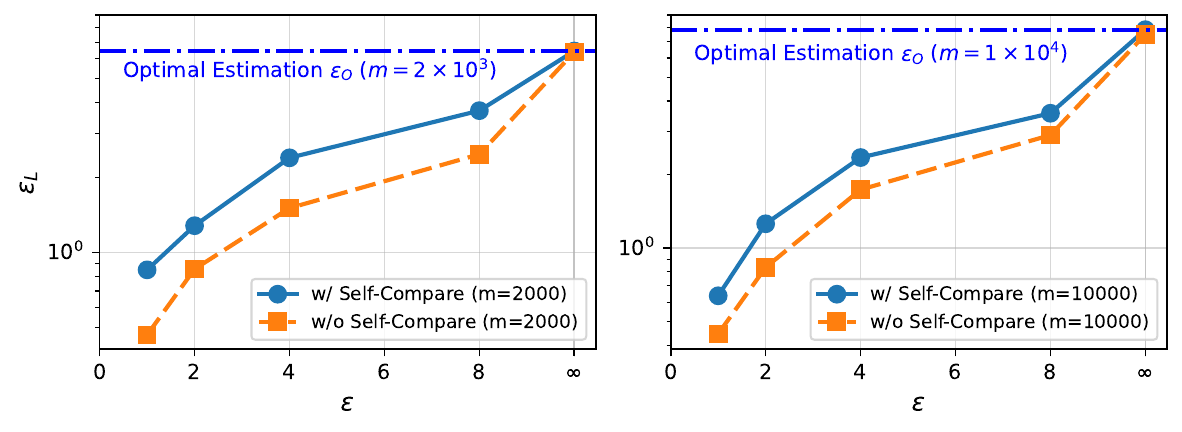}
    \caption{Effectiveness of self-comparison with MLP model}
    \label{fig:ab_comp}
\end{figure}

\subsection{Evaluation of Data-Independent Auditing}
We first evaluate \ours~ for data-independent auditing in \textbf{Case I} where the criterion of auditing algorithm is the tighter $\epsilon\_L$ compared to the upper bound $\epsilon$.

\subsubsection{Auditing Improvement across Privacy Budget}
We demonstrate the estimated privacy risk lower bound $\epsilon_L$ across different analytical privacy budget $\epsilon=\{1, 2, 4, 8, \infty \}$ in \Cref{fig:main_inf_eps}.
The non-DP training with $\epsilon=\infty$ in the figure acts as an indicator of the capability of $f_\text{MIA}$ in the auditing algorithm $\mathcal{A}$.
Given the optimal estimated risk $\epsilon_O$ calculated when we assume the MIA component $f_\text{MIA}$ has 100\% inference accuracy, thus if an auditing algorithm has $\epsilon_L<\epsilon_O$ for $\epsilon=\infty$, it indicates that the MIA $f_\text{MIA}$ is weak and impairs the auditing tightness.
We use orthogonal canaries with $m=2,000$ for ours, and optimize $m=\{500, 2,000, 5,000\}$ for each baselines of real-data canaries.

First of all, we can observe that the tightest $\epsilon_L$ (shown in gray shadow) across different $\epsilon$ is achieved with our \Cref{alg:canary} and the default architecture of 2-layer MLP.
The auditing performance on GPT-2 also approaches to the tightest results, while the gap between ours and all variants of \{\textit{In-distribution, Mislabeled, Poisoned}\} $\times$ \{GTSRB, CIFAR10, CIFAR100\} is significant.
The improvement comes from two aspects: the uncorrelated canary construction $f_\text{CNR}$ in \Cref{alg:canary} and the MIA capability of $f_\text{MIA}$ in \Cref{alg:comp}.

We note that baselines for image tasks have larger gap between $\epsilon_L$ and optimal estimates $\epsilon_O$ when $\epsilon=\infty$, which reflects the influence of data dependency on $f_\text{SCORE}\circ f_\text{MIA}$.
While \text{Mislabeled} baseline in LM task has higher $\epsilon_L$ than ours, the extra randomness introduced in DP training algorithm significantly limits its power when $\epsilon\neq \infty$.

\begin{table}[]
\centering
\caption{Estimated privacy lower bound $\epsilon_L$ under confidence $1-\beta=0.95$. The column with $\epsilon=\infty$ denotes the MIA capability limit in auditing.
	The optimal estimation $\epsilon_O$ is derived by assuming all predictions are correct, indicating the limit of statistic power limit in auditing.
}\label{tab:main_cv}
\resizebox{\linewidth}{!}{
\begin{tabular}{cc|ccc|ccc}
\toprule[1pt]
Dataset                                    & Model               & \multicolumn{3}{c}{\text{\(m = 2\times 10^3\) (\(\epsilon_O = 6.45\))}} & \multicolumn{3}{c}{\text{\(m = 10\times 10^3\) (\(\epsilon_O=7.83\))}} \\
\hline
\multicolumn{2}{c}{Indistribution}                               & $\epsilon=1$        & $\epsilon=8$        & $\epsilon=\infty$        & $\epsilon=1$         & $\epsilon=8$        & $\epsilon=\infty$        \\
\hline
CIFAR10                                    & CNN                 & 0.132        & 0.193        & 1.631          & 0.090         & 0.134        & 0.361          \\
                                           & ViT-Small           & 0.173        & 0.099        & 0.948          & 0.000         & 1.499        & 0.510          \\
                                           & Vit-Base            & 0.067        & 0.153        & 0.016          & 0.010         & 0.029        & 0.327          \\
CIFAR100                                   & CNN                 & 0.183        & 0.165        & 4.507          & 0.230         & 0.118        & 6.601          \\
                                           & ViT-Small           & 0.322        & 0.359        & 4.701          & 0.031         & 0.477        & 6.242          \\
                                           & Vit-Base            & 0.265        & 0.656        & 1.263          & 0.068         & 0.457        & 1.755          \\
GTSRB                                      & CNN                 & 0.004        & 0.131        & 1.986          & 0.003         & 0.105        & 0.368          \\
                                           & ViT-Small           & 0.101        & 0.387        & 0.479          & 0.091         & 0.050        & 3.534          \\
                                           & Vit-Base            & 0.376        & 0.981        & 1.189          & 0.053         & 0.061        & 0.259          \\
\hline
\multicolumn{2}{c}{Mislabeled}                                   & $\epsilon=1$        & $\epsilon=8$        & $\epsilon=\infty$        & $\epsilon=1$         & $\epsilon=8$        & $\epsilon=\infty$        \\
\hline
CIFAR10                                    & CNN                 & 0.739        & 1.219        & 1.730          & 0.068         & 0.465        & 1.168          \\
                                           & ViT-Small           & 0.036        & 0.099        & 1.703          & 0.151         & 0.655        & 1.793          \\
                                           & Vit-Base            & 0.001        & 0.739        & 0.196          & 0.110         & 1.195        & 0.928          \\
CIFAR100                                   & CNN                 & 0.038        & 0.291        & 0.286          & 0.068         & 0.052        & 0.047          \\
                                           & ViT-Small           & 0.228        & 0.708        & 2.881          & 0.084         & 0.641        & 1.722          \\
                                           & Vit-Base            & 0.310        & 0.745        & 2.571          & 0.059         & 0.884        & 1.215          \\
GTSRB                                      & CNN                 & 0.000        & 0.207        & 0.021          & 0.000         & 0.107        & 0.000          \\
                                           & ViT-Small           & 0.130        & 0.316        & 0.851          & 0.108         & 0.521        & 4.015          \\
                                           & Vit-Base            & 0.024        & 0.749        & 0.409          & 0.492         & 0.630        & 0.856          \\
                                           \hline
\multicolumn{2}{c}{Poisoned}                                     & $\epsilon=1$        & $\epsilon=8$        & $\epsilon=\infty$        & $\epsilon=1$         & $\epsilon=8$        & $\epsilon=\infty$        \\
\hline
CIFAR10                                    & CNN                 & 0.052        & 0.834        & 0.421          & 0.002         & 0.466        & 0.554          \\
                                           & ViT-Small           & 0.341        & 0.483        & 0.488          & 0.466         & 1.499        & 4.963          \\
                                           & Vit-Base            & 0.981        & 0.411        & 1.644          & 0.047         & 0.678        & 3.047          \\
CIFAR100                                   & CNN                 & 0.010        & 1.189        & 5.871          & 0.128         & 1.145        & 6.571          \\
                                           & ViT-Small           & 0.003        & 1.382        & 4.932          & 0.466         & 1.745        & 5.928          \\
                                           & Vit-Base            & 0.388        & 1.741        & 5.009          & 0.243         & 0.813        & 5.926          \\
GTSRB                                      & CNN                 & 0.523        & 1.124        & 3.506          & 0.033         & 0.958        & 0.976          \\
                                           & ViT-Small           & 0.000        & 0.153        & 2.338          & 0.000         & 0.993        & 2.655          \\
                                           & Vit-Base            & 0.000        & 0.158        & 2.414          & 0.580         & 0.595        & 0.958          \\
                                           \hline
\multicolumn{2}{c}{Ours}                         & $\epsilon=1$        & $\epsilon=8$        & $\epsilon=\infty$        & $\epsilon=1$         & $\epsilon=8$        & $\epsilon=\infty$        \\
\hline
\multirow{4}{*}{Gaussian w/o comp} & 2-Layer ReLU        & 0.322        & 2.362        & 6.395          & 0.207         & 2.096        & 7.816          \\
                                           & CNN          & 0.415        & 1.281        & 4.931          & 0.120             & 1.036            & 6.259              \\
                                           & ViT-Small           & 0.124        & 1.116        & 6.395          & 0.378             & 1.318            & 7.503              \\
                                           & Vit-Base            & 0.238        & 1.038        & 6.395          & 0.244             & 1.422            & 7.503              \\
                                           \hline
\multirow{4}{*}{Gaussian w/ comp}  & 2-Layer ReLU        & 0.620        & 2.707        & 6.449          & \textbf{0.771}         & \textbf{3.780}        & 7.829          \\
                                           & CNN           & 0.579        & 1.281        & 5.395          & 0.171         & 1.423        & 6.816          \\
                                           & ViT-Small           & 0.413        & 1.306        & 6.449          & 0.173         & 1.446        & 7.834          \\
                                           & Vit-Base            & 0.174        & 1.365        & 6.449          & 0.521         & 1.601        & 7.834          \\
                                           \hline
Orthogonal w/o comp               & 2-Layer ReLU & 0.447        & 3.032        & 6.395          & 0.501             & 3.191            & 7.503              \\
Orthogonal w/ comp                & 2-Layer ReLU & \textbf{1.089}        & \textbf{3.059}        & \textbf{6.449}          & 0.623         & 3.270        & \textbf{7.834}         \\
\bottomrule[1pt]
\end{tabular}
}
\end{table}

Additionally, we set the same training hyper-parameters for all baselines and our auditing for a fair comparison.
Thus, our improvement does not depend on model architecture, even lower than MLP model and GPT-2 model, $\epsilon_L$ for CNN still outperforms other baselines with real-data as canary and sorting for MIA.
Specifically for each baseline in each dataset and each model combination, we demonstrate a detailed result in \Cref{tab:main_cv}.
We can observe that 

\subsubsection{Auditing Improvement across Different $m$}
Now we compare \ours~ with baselines across $m$, the number of auditing data.
The size of audit data is vital to the auditing performance, as it not only controls the statistical power as reflected in $\epsilon_O$, and also influences the dependency level for the $m$ observations.
The larger $m$ results in larger $\epsilon_O$ but may also involves more dependency that hurt the auditing power.

As shown in \Cref{fig:inf_m}, the two types of synthetic canary that we propose in \Cref{alg:canary} achieve a better auditing result than all baseline, as a result of uncorrelated audit data and the self-comparison framework.
By comparing different baselines, we find that \textit{Mislabeled} is superior than other baselines, due to its less dependency on label space, but still inferior to our methods with both noisy features and noisy labels because noisy features are more separated in the space~\cite{zhang2016understanding}.

In the right sub-figure for language modeling task, we compare with the strongest baseline \textit{NewToken}.
For a fair comparison, we insert the same number of new tokens, keep all training hyper-parameters the same, e.g., using the same supervised fine-tuning loss by masking the loss calculation in the prefix.
The key difference between ours and \textit{NewToken} is that they use random prefix and suffix while ours use repeated tokens as prefix and suffix to construct the canary.
Our implementation has achieved the their reported optimal $\epsilon_L\approx 1.3$ given $\epsilon=4, \delta=1e-5$ given 95\% confidence.
And \ours~ is superior than theirs across different $m=n$.

In general, for both tasks, $\epsilon_L$ increases and then decreases when $m$ gets larger for all methods, which reflecting the expected trade-off controlled by $m$.
The optimal choice of $m$ for maximizing auditing performance is around $5\times 10^2-2\times10^3$ across tasks in case I with $n=m$.

\subsubsection{Effectiveness of Orthogonal Features}
As shown in \Cref{fig:inf_m} and \Cref{tab:main_cv}, orthogonal features outperforms gaussian features across different audit data size, especially for a relatively small $m$ or $n$. 
The critical point is attributed to the feature dimension that we set $d_x=10^3$.
Because the feature vectors of our orthogonal variant become approximated orthogonal when $m>d_x$, which explains the drop of $\epsilon_L$ in the left sub-figure for $m>10^3$.
Thus, it is expected that a larger $d_x$ would result in a larger critical point of $m$ for our orthogonal method.

\subsubsection{Effectiveness of Self-Comparison}
Above auditing improvement comes from our canary construction with uncorrelated data and the self-comparison framework.
It should be noted that the self-comparison cannot stand without the uncorrelated canaries.
So we now ablate effectiveness of self-comparison for further analysis.

As shown in \Cref{fig:ab_comp}, for both $m=2\times 10^3$ and $m=1\times10^4$, self-comparison significantly outperforms than the version without it, even with the same canary construction.
By comparing the version w/o self-comparison with real-data best result (e.g., $\epsilon_L\approx 1.3$ for $\epsilon=4$~\cite{steinke2023privacy, panda2025privacy}), it shows a slight improvement due to the pure uncorrelated canary sample.
This shows that the independent scoring $f_\text{SCORE}$ and MIA decision $f_\text{MIA}$ matters.
While existing works on improving the canary only~\cite{panda2025privacy, nasr2021adversary, steinke2023privacy}, \ours~ improves more by consider to reduce the data dependency in \textbf{both} training and MIA inference.

\begin{figure}
\centering
\includegraphics[width=\linewidth]{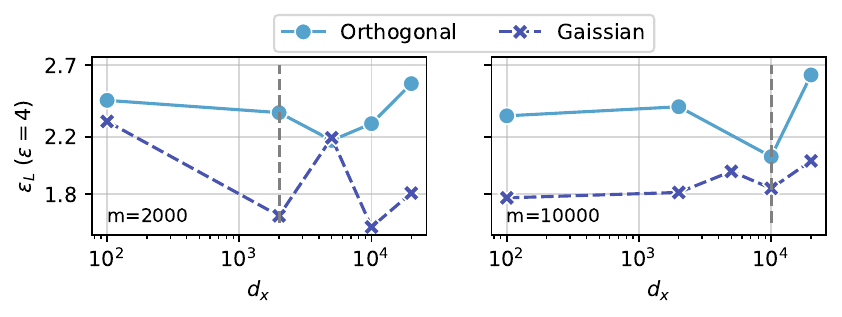}
\caption{Influence of feature dimension $d_x$}
\label{fig:inf_dx_cv}
\end{figure}

\subsubsection{Influence of Canary Dimensions}
As observed in \Cref{fig:inf_m}, the input feature dimension matters especially for our orthogonal variant.
Thus, we deeply analyze its influence in \Cref{fig:inf_dx_cv}.
For two different $m \in \{2\times 10^3, 1\times10^4\}$, we can observe that the estimated $\epsilon_L$ first decreases then increases with larger $d_x$ after a critical point.
The decrease is not monotonous and not as significant as the increase after the critical point.
According to the gray dashed vertical line, it shows that the critical point comes after the $d_x\ge m$ (with $m=n$) for both our Gaussian and Orthogonal variants.
Thus, it suggests that a safe choice of the dimension of input features $d_x \gg m$, aligning with a similar conclusion drawn in a recent auditing study~\cite{xiang2025privacy}.

\begin{figure}
\centering
\includegraphics[width=\linewidth]{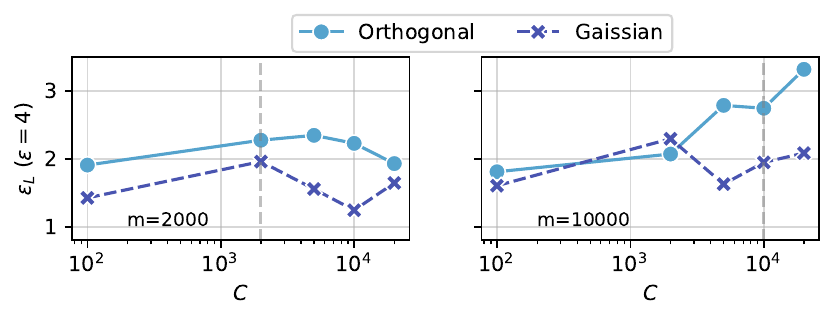}
    \caption{Influence of the number of target encoding space $C$.}
    \label{fig:inf_c}
\end{figure}

\begin{figure}
\centering
\includegraphics[width=0.8\linewidth]{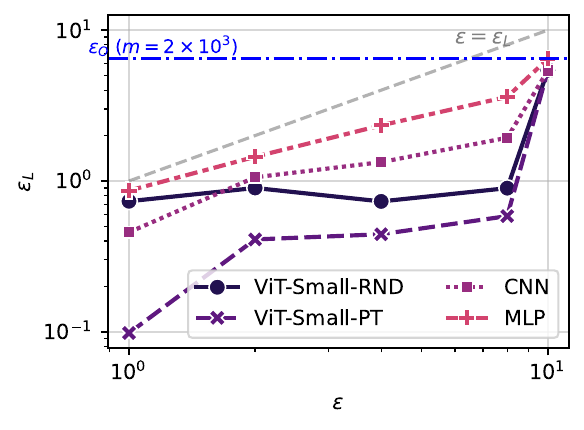}
    \caption{Influence of the number of target encoding space $C$.}
    \label{fig:inf_model_cv}
\end{figure}

\subsubsection{Influence of Encoding Space}
We find that the number of encoding space size $C$ is vital hyper-parameters, because it depends how many $m$ can be sampled uniquely, or in other words, how sparsity can the membership encoding space $\mathcal{X}\times\mathcal{Y}$ (or terms of $\mathcal{G}\times\mathcal{E}$ in Case II) is.
In \Cref{fig:inf_c}, it is interesting to find that the optimal choice of $C$ emerges after the critical point of $C=m$.
Intuitively, it indicates that unique encoding is important for a good membership encoding performance.

While for relatively smaller $m$ (e.g., $2\times 10^3$) the $\epsilon_L$ slightly crease after the optimal $C$, $\epsilon_L$ continues to increase for larger $m=1\times10^4$.
The trade-off here is that the number of model parameters scales up with larger $C$, and the signal to noise ratio in DP training decreases with small $m$, which makes a larger model harder to converge.
But it is not a issue when $m$ is large enough (e.g., $m=1\times 10^4$), otherwise it can be solved by setting an even larger sub-sampling ratio in DP implementation.

\subsubsection{Influence of Pre-Training.}
It should be noted that our \ours~ is independent of model architecture by design.
While we assume auditor cannot craft the model initialization as \textit{Poisoned}, the auditor can require the specific model architecture.
For Case I where maxing auditing is the only criterion, it is meaningful to discuss what model should auditor 
designate for stronger auditing.

In \Cref{fig:inf_model_cv}, we compare variants of \ours~ on different model architectures, such as CNN with $8\times 10^5$, MLP with $3\times 10^7$ and $2.2 \times 10^7$ parameters.
Obviously, 2-layer MLP model with ReLU activations achieves the strongest auditing performance, probably because the large hidden states dimension $d_h=10^5$.
But the auditing strength does not simply scale with the trainable parameter size, we also notice that ViT-Small results in smaller $\epsilon_L$.
Comparing with training from scratch ViT-Small-RND with ViT-Small-PT, we find that random initialization benefits more for our synthetic canaries, especially for a small analytical privacy budget.
This is because the mode initialization pre-trained on real-world dataset already finds a relatively flatten loss landscape, and it is more robust to noise.
Otherwise, it requires more iterations for the pre-training model to fit the sample-specific noise in our canary.
Thus, we suggest to use a random initialized neural network (such as 2-layer MLP with ReLU activations) in \ours~ to maximize the auditing power.

\subsection{Evaluation of Data-Dependent Auditing}
In this section, we perform auditing for \textbf{Case II}. where both good model utility and auditing capability are criterion.
For the baseline methods to achieve the best utility and auditing trade-off, we use the optimal hyper-parameters such as learning rate, batch size that we grid-searched for in \Cref{fig:demo_challenge1}.

\begin{figure}
	\centering
    \includegraphics[width=\linewidth]{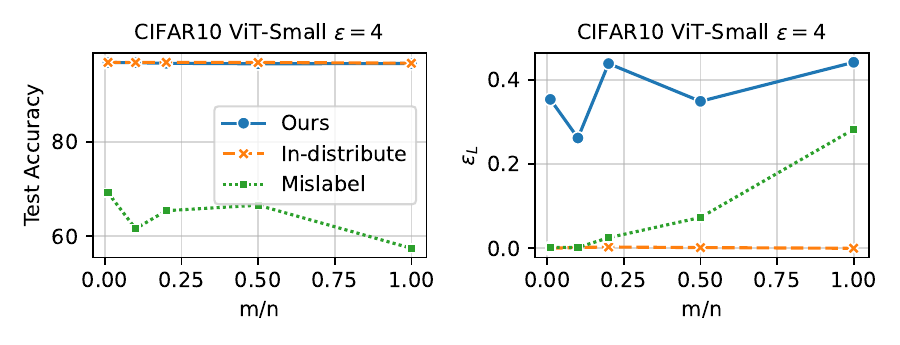}
	\includegraphics[width=\linewidth]{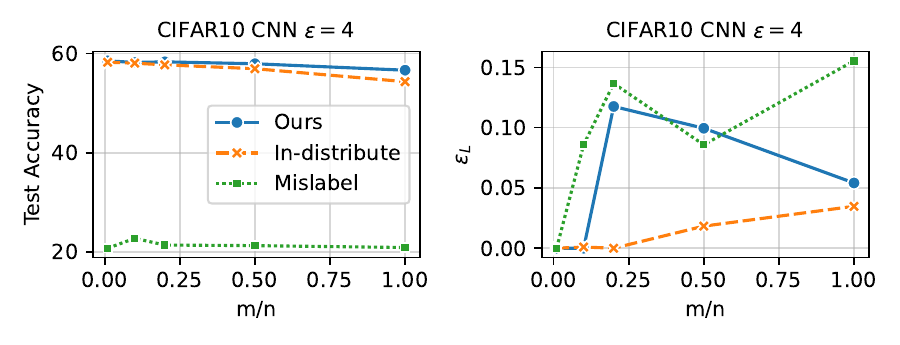}
	\includegraphics[width=0.49\linewidth]{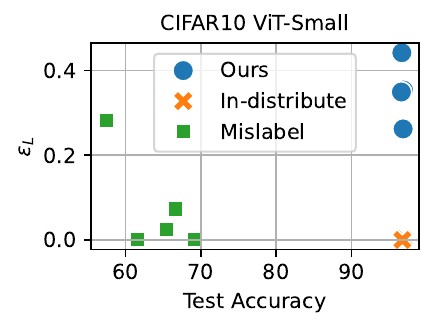}
	\includegraphics[width=0.49\linewidth]{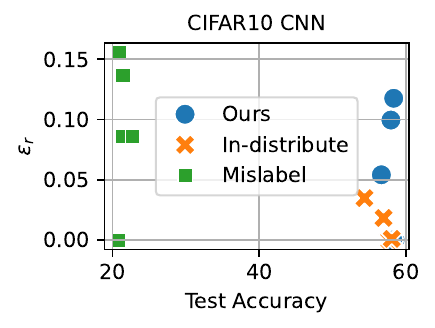}
    \caption{Utility and Auditing Trade-off in Case II for Image Classification}
    \label{fig:case3_cv}
\end{figure}

\begin{figure}
\centering
\includegraphics[width=\linewidth]{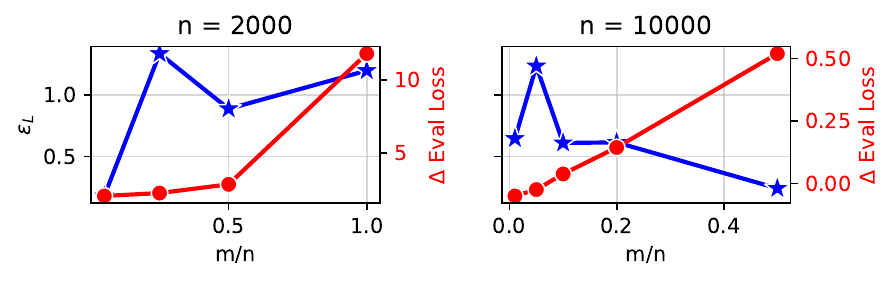}
\caption{Utility and Auditing Trade-off in Case II for language modeling on GPT-2.
The $\Delta$ evaluation loss indicates the loss difference of \ourss~ with canaries minus the \textit{In-distribution} baseline with all normal samples.
}
\label{fig:case3_lm}
\end{figure}

\subsubsection{Trade-off for Image Classification}
As shown in \Cref{fig:case3_cv}, we evaluate with two representative models: training from scratch for CNN and fine-tuning for pre-trained ViT-Small.
We use a special instance of our random canary by repeating a randomly sampled pixel from a \( 3 \times 256 \times 256 \) image, and apply it as a patch of dimension \( d_x = 3 \times 10 \times 10 \) to a real-world image.
To reduce the required weights for the auditing head, we set $|\mathcal{E}|=100$ in \Cref{alg:multi_task_data}.

It is clear that for both models, our method achieves the best utility-auditing trade-off than other O(1) variants.
For pre-trained models such as ViT, we notice that \ourss~ achieves the best of the word: near zero utility loss compared to O(1) with original samples as canary; while large improvement on $\epsilon_L$.
While \textit{Mislabeled} has better $\epsilon_L$ compared to \textit{In-Distribution}, it significantly hurts the utility even when the ratio of mislabel $m/n$ is small.
The success of \ourss~ is attributed to our insight in \Cref{sec:insight} of decoupling training objective, which disentangles the implicit conflict between utility and auditing.

It is also interesting to notice that for CNN models trained from random initialization, \ourss~ can even outperform \textit{In-Distribution} baseline that use normal samples as canary, while our $\epsilon_L$ is comparable to another baseline \textit{Mislabeled} which absolutely abandons the utility.
The key reason of why we have better utility compared \textit{In-Distribution} is attributed to our \Cref{alg:comp} which leverages $m$ samples in training.
By contrast, even though the canaries in \textit{In-Distribution} are clean samples, the Bernoulli sampling in the original O(1) framework results in a loss of around $m/2$ normal samples for improving utility.
And the loss on the amount of training data scales up with a larger $m/n$, as shown in \Cref{fig:case3_cv}.

We do not consider comparison with \textit{Poisoned} in case II because it requires the auditor to have a public in-distribution auxiliary dataset and craft the model initialization by warming up on the auxiliary dataset.
Additionally, it is unfair to compare a pure private-preserving training with public-assisted training.

\subsubsection{Trade-off for Language Modeling}
We also analyze the utility and privacy trade-off in language modeling for case II.
It should be noted that there is a distinguished difference between image classification task and language tasks in case II.
In image classification task, the utility and auditing power conflicts because audit sample's label share the same space as the normal non-audit samples.
As an example, \textit{Mislabeled} crafts the true label while still shares the space $\mathcal{Y}$.
While for language models, when the training sample only includes augmented new tokens, although parameters of intermediate layers might change, the normal token embeddings do not change.
Thus, as shown in \Cref{fig:case3_lm}, when the total amount of training data $n$ is relatively large, the utility degradation compared to \textit{In-Distribution} with all normal samples is trivial.

In general, we observe a consistent trend for both $n=\{2\times 10^3, 1\times 10^4\}$ that the auditing capability (i.e., $\epsilon_L$) first increases then decreases.
The increase is caused with a higher statistical power limit on $\epsilon_O$; while the decrease is attributed to the degradation on the membership encoding sparsity which  roughly scale according to $m/\sqrt{|V_\text{new}|}$.

\section{Discussion and Broader Impacts}
Towards practical auditing tools, our work builds on the efficient O(1) auditing framework under a realistic black-box access assumption. Although it is well known that combining these two settings makes auditing more challenging~\cite{steinke2023privacy,nasr2021adversary}, we believe it is necessary to position DP auditing as a seamlessly integrated component of machine learning pipeline -- strengthening data governance and enabling continuous privacy monitoring, especially in sensitive domains such as healthcare analytics and government AI systems where theoretical guarantees demand empirical validation.

Unlike prior efforts that tighten theoretical auditing bounds via alternative confidence intervals~\cite{lu2022general,zeng2020meddialog} or novel DP definitions~\cite{mahloujifar2024auditing,xiang2025privacy}, our study offers fresh insights into black-box O(1) DP auditing by systematically reframing \textit{the goal of DP auditing} from the perspective of \emph{efficiency-utility-auditing} trade-off space-—a critical dimension that has been largely overlooked~\cite{jagielski2020auditing,nasr2023tight,steinke2023privacy}.

By revealing the fundamental challenge of O(1) auditing, we categorize the auditing goals as:
1) Data-independent auditing, which targets the DP implementation itself; 
2) Data-dependent auditing, which evaluates the empirical privacy risk of a specific training run on a given dataset and model.
Our problem formulation opens two pivotal research directions: 
\begin{enumerate}
    \item Development of better membership encoding tasks that can maximize auditing tightness for the data-independent DP implementation.
    It is critical for monitoring the commercial Machine-Learning-as-a-Service compliance to privacy guidelines.
    More specifically, based on our formulation of membership encoding loss, future works are encouraged to design better canary data construction and model architecture for facilitating a tighter auditing.
    \item Co-design of non-intrusive audit sub-tasks that intrinsically enhance membership signals \emph{during} primary model training without hurting main task utility.
    This auditing-along-training framework is useful especially for the scenario when data-dependent risk is the target or when a single DP training is expensive.
    Based on our key insight of decoupling training objective, better auditing sub-task helps to further mitigate the fundamental conflict of utility and auditing.
\end{enumerate} 

Our methodology bridges the critical gap between formal privacy analysis and practical deployment constraints, establishing new pipelines for trustworthy data management in the widely available black-box learning scenarios.
While we evaluate on image classification and language modeling, we anticipate extending both auditing frameworks to other tasks—such as image generation—via tailored membership encoding strategies and sub-task designs.

\section{Conclusions}
In this work, we rethink the fundamental challenge of black-box O(1) auditing and formulate two problems: data-independent and data-dependent auditing.
We identify two bottlenecks: inevitable data dependency and implicit conflict between utility and auditing, which either remain as open problems or are ignored by existing works.
To mitigate data dependency, we propose \ours~ with a novel canary construction using uncorrelated features and labels, formulating auditing as a membership encoding task with a pair-matching objective. 
To further reduce dependency among audit samples during the membership inference stage, we introduce a self-comparison framework.
We extend \ours~ to \ourss~ for data-dependent scenarios, where audit results are estimated alongside models with acceptable utility, framing it as a novel multi-task problem by decoupling training objectives. 
Evaluation across image classification and language modeling tasks shows that our black-box O(1) data-independent auditing outperforms state-of-the-art approaches while saving thousands of training runs, and the data-dependent auditing achieves a better auditing-utility trade-off.

\IEEEtriggercmd{\enlargethispage{-5in}}

\bibliographystyle{IEEEtran}
\bibliography{bib/dp,bib/ml,bib/poison,bib/crypto,bib/llm,bib/ppml,bib/attack}

\begin{thebibliography}{10}
\providecommand{\url}[1]{#1}
\csname url@samestyle\endcsname
\providecommand{\newblock}{\relax}
\providecommand{\bibinfo}[2]{#2}
\providecommand{\BIBentrySTDinterwordspacing}{\spaceskip=0pt\relax}
\providecommand{\BIBentryALTinterwordstretchfactor}{4}
\providecommand{\BIBentryALTinterwordspacing}{\spaceskip=\fontdimen2\font plus
\BIBentryALTinterwordstretchfactor\fontdimen3\font minus \fontdimen4\font\relax}
\providecommand{\BIBforeignlanguage}[2]{{%
\expandafter\ifx\csname l@#1\endcsname\relax
\typeout{** WARNING: IEEEtran.bst: No hyphenation pattern has been}%
\typeout{** loaded for the language `#1'. Using the pattern for}%
\typeout{** the default language instead.}%
\else
\language=\csname l@#1\endcsname
\fi
#2}}
\providecommand{\BIBdecl}{\relax}
\BIBdecl

\bibitem{zhu2019deep}
L.~Zhu, Z.~Liu, and S.~Han, ``Deep leakage from gradients,'' \emph{Advances in neural information processing systems}, vol.~32, 2019.

\bibitem{carlini2022membership}
N.~Carlini, S.~Chien, M.~Nasr, S.~Song, A.~Terzis, and F.~Tramer, ``Membership inference attacks from first principles,'' in \emph{2022 IEEE Symposium on Security and Privacy (SP)}.\hskip 1em plus 0.5em minus 0.4em\relax IEEE, 2022, pp. 1897--1914.

\bibitem{carlini2021extracting}
N.~Carlini, F.~Tramer, E.~Wallace, M.~Jagielski, A.~Herbert-Voss, K.~Lee, A.~Roberts, T.~Brown, D.~Song, U.~Erlingsson \emph{et~al.}, ``Extracting training data from large language models,'' in \emph{30th USENIX Security Symposium (USENIX Security 21)}, 2021, pp. 2633--2650.

\bibitem{abadi2016deep}
M.~Abadi, A.~Chu, I.~Goodfellow, H.~B. McMahan, I.~Mironov, K.~Talwar, and L.~Zhang, ``Deep learning with differential privacy,'' in \emph{Proceedings of the 2016 ACM SIGSAC Conference on Computer and Communications Security}, 2016, pp. 308--318.

\bibitem{dwork2016calibrating}
C.~Dwork, F.~McSherry, K.~Nissim, and A.~Smith, ``Calibrating noise to sensitivity in private data analysis,'' \emph{Journal of Privacy and Confidentiality}, vol.~7, no.~3, pp. 17--51, 2016.

\bibitem{shokri2017membership}
R.~Shokri, M.~Stronati, C.~Song, and V.~Shmatikov, ``Membership inference attacks against machine learning models,'' in \emph{2017 IEEE symposium on security and privacy (SP)}.\hskip 1em plus 0.5em minus 0.4em\relax IEEE, 2017, pp. 3--18.

\bibitem{tramer2022debugging}
F.~Tramer, A.~Terzis, T.~Steinke, S.~Song, M.~Jagielski, and N.~Carlini, ``Debugging differential privacy: A case study for privacy auditing,'' \emph{arXiv preprint arXiv:2202.12219}, 2022.

\bibitem{jagielski2020auditing}
M.~Jagielski, J.~Ullman, and A.~Oprea, ``Auditing differentially private machine learning: How private is private sgd?'' \emph{Advances in Neural Information Processing Systems}, vol.~33, pp. 22\,205--22\,216, 2020.

\bibitem{nasr2021adversary}
M.~Nasr, S.~Songi, A.~Thakurta, N.~Papernot, and N.~Carlin, ``Adversary instantiation: Lower bounds for differentially private machine learning,'' in \emph{2021 IEEE Symposium on security and privacy (SP)}.\hskip 1em plus 0.5em minus 0.4em\relax IEEE, 2021, pp. 866--882.

\bibitem{nasr2023tight}
M.~Nasr, J.~Hayes, T.~Steinke, B.~Balle, F.~Tram{\`e}r, M.~Jagielski, N.~Carlini, and A.~Terzis, ``Tight auditing of differentially private machine learning,'' in \emph{32nd USENIX Security Symposium (USENIX Security 23)}, 2023, pp. 1631--1648.

\bibitem{steinke2023privacy}
T.~Steinke, M.~Nasr, and M.~Jagielski, ``Privacy auditing with one (1) training run,'' \emph{Advances in Neural Information Processing Systems}, vol.~36, pp. 49\,268--49\,280, 2023.

\bibitem{xiang2025privacy}
Z.~Xiang, T.~Wang, and D.~Wang, ``Privacy audit as bits transmission:(im) possibilities for audit by one run,'' \emph{arXiv preprint arXiv:2501.17750}, 2025.

\bibitem{mahloujifar2024auditing}
S.~Mahloujifar, L.~Melis, and K.~Chaudhuri, ``Auditing $ f $-differential privacy in one run,'' \emph{arXiv preprint arXiv:2410.22235}, 2024.

\bibitem{muthu2024nearly}
M.~S. Muthu Selva~Annamalai and E.~De~Cristofaro, ``Nearly tight black-box auditing of differentially private machine learning,'' \emph{Advances in Neural Information Processing Systems}, vol.~37, pp. 131\,482--131\,502, 2024.

\bibitem{panda2023differentially}
A.~Panda, T.~Wu, J.~Wang, and P.~Mittal, ``Differentially private in-context learning,'' in \emph{The 61st Annual Meeting of the Association for Computational Linguistics}, 2023.

\bibitem{jayaraman2019evaluating}
B.~Jayaraman and D.~Evans, ``Evaluating differentially private machine learning in practice,'' in \emph{28th USENIX Security Symposium (USENIX Security 19)}, 2019, pp. 1895--1912.

\bibitem{maddock2022canife}
S.~Maddock, A.~Sablayrolles, and P.~Stock, ``Canife: Crafting canaries for empirical privacy measurement in federated learning,'' \emph{arXiv preprint arXiv:2210.02912}, 2022.

\bibitem{lu2022general}
F.~Lu, J.~Munoz, M.~Fuchs, T.~LeBlond, E.~Zaresky-Williams, E.~Raff, F.~Ferraro, and B.~Testa, ``A general framework for auditing differentially private machine learning,'' \emph{Advances in Neural Information Processing Systems}, vol.~35, pp. 4165--4176, 2022.

\bibitem{zanella2023bayesian}
S.~Zanella-B{\'e}guelin, L.~Wutschitz, S.~Tople, A.~Salem, V.~R{\"u}hle, A.~Paverd, M.~Naseri, B.~K{\"o}pf, and D.~Jones, ``Bayesian estimation of differential privacy,'' in \emph{International Conference on Machine Learning}.\hskip 1em plus 0.5em minus 0.4em\relax PMLR, 2023, pp. 40\,624--40\,636.

\bibitem{andrew2023one}
G.~Andrew, P.~Kairouz, S.~Oh, A.~Oprea, H.~B. McMahan, and V.~M. Suriyakumar, ``One-shot empirical privacy estimation for federated learning,'' \emph{arXiv preprint arXiv:2302.03098}, 2023.

\bibitem{pillutla2023unleashing}
K.~Pillutla, G.~Andrew, P.~Kairouz, H.~B. McMahan, A.~Oprea, and S.~Oh, ``Unleashing the power of randomization in auditing differentially private ml,'' \emph{Advances in Neural Information Processing Systems}, vol.~36, pp. 66\,201--66\,238, 2023.

\bibitem{liu2024precurious}
R.~Liu, T.~Wang, Y.~Cao, and L.~Xiong, ``Precurious: How innocent pre-trained language models turn into privacy traps,'' in \emph{Proceedings of the 2024 ACM SIGSAC Conference on Computer and Communications Security}, 2024.

\bibitem{wen2024privacy}
Y.~Wen, L.~Marchyok, S.~Hong, J.~Geiping, T.~Goldstein, and N.~Carlini, ``Privacy backdoors: Enhancing membership inference through poisoning pre-trained models,'' \emph{arXiv preprint arXiv:2404.01231}, 2024.

\bibitem{huang2024general}
Z.~Huang, N.~Z. Gong, and M.~K. Reiter, ``A general framework for data-use auditing of ml models,'' in \emph{Proceedings of the 2024 on ACM SIGSAC Conference on Computer and Communications Security}, 2024, pp. 1300--1314.

\bibitem{tongmuch}
Y.~Tong, J.~Ye, S.~Zarifzadeh, and R.~Shokri, ``How much of my dataset did you use? quantitative data usage inference in machine learning,'' in \emph{The Thirteenth International Conference on Learning Representations}.

\bibitem{song2020membership}
C.~Song and R.~Shokri, ``Membership encoding for deep learning,'' in \emph{Proceedings of the 15th ACM Asia Conference on Computer and Communications Security}, 2020, pp. 344--356.

\bibitem{chen2024method}
Z.~Chen and K.~Pattabiraman, ``A method to facilitate membership inference attacks in deep learning models,'' \emph{arXiv preprint arXiv:2407.01919}, 2024.

\bibitem{kairouz2015composition}
P.~Kairouz, S.~Oh, and P.~Viswanath, ``The composition theorem for differential privacy,'' in \emph{International conference on machine learning}.\hskip 1em plus 0.5em minus 0.4em\relax PMLR, 2015, pp. 1376--1385.

\bibitem{panda2025privacy}
A.~Panda, X.~Tang, M.~Nasr, C.~A. Choquette-Choo, and P.~Mittal, ``Privacy auditing of large language models,'' \emph{arXiv preprint arXiv:2503.06808}, 2025.

\bibitem{wu2020generalization}
S.~Wu, H.~Zhang, G.~Valiant, and C.~R{\'e}, ``On the generalization effects of linear transformations in data augmentation,'' in \emph{International conference on machine learning}.\hskip 1em plus 0.5em minus 0.4em\relax PMLR, 2020, pp. 10\,410--10\,420.

\bibitem{zhang2016understanding}
C.~Zhang, S.~Bengio, M.~Hardt, B.~Recht, and O.~Vinyals, ``Understanding deep learning requires rethinking generalization,'' \emph{arXiv preprint arXiv:1611.03530}, 2016.

\bibitem{bellare2001introduction}
M.~Bellare and P.~Rogaway, ``Introduction to modern cryptography,'' \emph{Lecture Notes}, 2001.

\bibitem{panda2024new}
A.~Panda, X.~Tang, S.~Mahloujifar, V.~Sehwag, and P.~Mittal, ``A new linear scaling rule for private adaptive hyperparameter optimization,'' in \emph{International Conference on Machine Learning}.\hskip 1em plus 0.5em minus 0.4em\relax PMLR, 2024, pp. 39\,364--39\,399.

\bibitem{papernot2021tempered}
N.~Papernot, A.~Thakurta, S.~Song, S.~Chien, and {\'U}.~Erlingsson, ``Tempered sigmoid activations for deep learning with differential privacy,'' in \emph{Proceedings of the AAAI Conference on Artificial Intelligence}, vol.~35, no.~10, 2021, pp. 9312--9321.

\bibitem{dormann2021not}
F.~D{\"o}rmann, O.~Frisk, L.~N. Andersen, and C.~F. Pedersen, ``Not all noise is accounted equally: How differentially private learning benefits from large sampling rates,'' in \emph{2021 IEEE 31st International Workshop on Machine Learning for Signal Processing (MLSP)}.\hskip 1em plus 0.5em minus 0.4em\relax IEEE, 2021, pp. 1--6.

\bibitem{bu2022automatic}
Z.~Bu, Y.-X. Wang, S.~Zha, and G.~Karypis, ``Automatic clipping: Differentially private deep learning made easier and stronger,'' \emph{Advances in Neural Information Processing Systems}, vol.~36, 2023.

\bibitem{de2022unlocking}
S.~De, L.~Berrada, J.~Hayes, S.~L. Smith, and B.~Balle, ``Unlocking high-accuracy differentially private image classification through scale,'' \emph{arXiv preprint arXiv:2204.13650}, 2022.

\bibitem{dosovitskiy2020image}
A.~Dosovitskiy, L.~Beyer, A.~Kolesnikov, D.~Weissenborn, X.~Zhai, T.~Unterthiner, M.~Dehghani, M.~Minderer, G.~Heigold, S.~Gelly \emph{et~al.}, ``An image is worth 16x16 words: Transformers for image recognition at scale,'' \emph{arXiv preprint arXiv:2010.11929}, 2020.

\bibitem{li2021large}
X.~Li, F.~Tramer, P.~Liang, and T.~Hashimoto, ``Large language models can be strong differentially private learners,'' \emph{arXiv preprint arXiv:2110.05679}, 2021.

\bibitem{yu2021differentially}
D.~Yu, S.~Naik, A.~Backurs, S.~Gopi, H.~A. Inan, G.~Kamath, J.~Kulkarni, Y.~T. Lee, A.~Manoel, L.~Wutschitz \emph{et~al.}, ``Differentially private fine-tuning of language models,'' \emph{arXiv preprint arXiv:2110.06500}, 2021.

\bibitem{zeng2020meddialog}
G.~Zeng, W.~Yang, Z.~Ju, Y.~Yang, S.~Wang, R.~Zhang, M.~Zhou, J.~Zeng, X.~Dong, R.~Zhang \emph{et~al.}, ``Meddialog: Large-scale medical dialogue datasets,'' in \emph{Proceedings of the 2020 Conference on Empirical Methods in Natural Language Processing (EMNLP)}, 2020.

\end{thebibliography}

\end{document}